\documentclass[11pt]{article}
\usepackage{jcappub}
\usepackage{mathtools}
\usepackage{bm}
\usepackage{dsfont}
\usepackage{color}
\usepackage{array}
\usepackage{graphicx}
\usepackage{subcaption}
\usepackage{soul}
\usepackage{multirow}
\usepackage{multicol}
\usepackage{float}
\usepackage{hhline}
\usepackage[dvipsnames]{xcolor}
\usepackage[normalem]{ulem}
\usepackage{mathrsfs}
\usepackage{wasysym}
\usepackage[mathscr]{euscript}
\usepackage{ulem}

\title{Gravitational Waves from Mergers of Asymmetric Dark Stars}

\author[a]{Boris Betancourt Kamenetskaia,}
\author[a]{Qianhang Ding,}
\author[b]{Chris Kouvaris}
\affiliation[a]{Cosmology, Gravity, and Astroparticle Physics Group, Center for Theoretical Physics of the Universe,
Institute for Basic Science (IBS), Daejeon, 34126, Korea}
\affiliation[b]{Physics Division, National Technical University of Athens, 15780 Zografou Campus, Athens, Greece}

\emailAdd{laybors@ibs.re.kr}
\emailAdd{dingqh@ibs.re.kr}
\emailAdd{kouvaris@mail.ntua.gr}

\abstract{A strongly self-interacting component of asymmetric dark matter (DM) particles can form compact dark stars (DSs). These objects have a broad spectrum of masses and radii, with distinct evolution histories from both neutron stars and black holes (BHs). We argue that these differences allow a population of DSs to contribute significantly to the astrophysical merger rate in unique and discernible ways. Specifically, their merger rate could dominate at low redshifts over other sources, while their mass function may populate windows outside known astrophysical processes. We investigate the structure and formation of DSs within a dissipative model, and calculate the enhancement of their merger cross-section due to tidal deformation effects. From this, we derive the present-day merger rate and its differential mass distribution. These findings open a new window to probe DM substructure and particle interactions through present and future gravitational wave (GW) observatories.
}

\begin{document}
\maketitle
\flushbottom

\section{Introduction}

Dark matter (DM) constitutes approximately 27\% of the energy budget of the Universe~\cite{Planck:2018vyg}, yet its fundamental nature remains one of the most pressing open questions in modern physics. Despite overwhelming evidence for its existence from galactic to cosmological scales, the properties of the DM are still unknown. If DM is in the form of particles, its mass, spin, and interaction strengths can span a vast parameter space. A central question in this context is whether DM is symmetric or asymmetric in nature. An intriguing possibility is that DM carries a conserved ``dark charge'' (analogous to baryon number), and that an asymmetry exists between DM particles and antiparticles in the early Universe, mirroring the baryon-antibaryon asymmetry in the visible sector~\cite{Nussinov:1985xr,Barr:1990ca,Gudnason:2006yj}. In this asymmetric DM scenario~\cite{Zurek:2013wia,Petraki:2013wwa}, DM antiparticles annihilate away, leaving behind a population of stable particles that cannot self-annihilate due to charge conservation.

If DM possesses strong self-interactions, a fraction of this asymmetric population could collapse gravitationally, leading to the formation of stable, compact objects known as dark stars (DSs)~\cite{Kouvaris:2015rea,Eby:2015hsq}.\footnote{This type of stars are in clear distinction to DSs made of symmetric DM where DM annihilations are present~\cite{Spolyar:2007qv}.} This process is analogous to the formation of ordinary stars from baryonic matter. The possibility of such gravitationally bound DM systems was first explored by Kaup~\cite{Kaup:1968zz}, who considered a configuration of non-interacting complex scalar fields supported against collapse by the Heisenberg uncertainty principle. This work was later extended to include the impact of self-interactions for bosonic fields~\cite{Colpi:1986ye} and for fermionic DM~\cite{Kouvaris:2015rea}. It has been argued that such compact objects could be formed dynamically through mechanisms like gravitational cooling~\cite{Seidel:1993zk} or the bremsstrahlung of dark photons~\cite{Chang:2018bgx}.

The observational search for these DSs employs a variety of techniques, with the most powerful constraints deriving from their gravitational interactions. Depending on their mass, DSs can be probed by stellar microlensing~\cite{Macho:2000nvd,EROS-2:2006ryy,Niikura:2019kqi}, supernovae magnification~\cite{Zumalacarregui:2017qqd}, gravitational waves (GWs) from dark compact objects~\cite{Maselli:2017vfi,Kavanagh:2018ggo,LIGOScientific:2019kan,Chen:2019irf}, and dynamical effects on wide binaries~\cite{Monroy-Rodriguez:2014ula} or dwarf galaxies~\cite{Brandt:2016aco}. A synthesis of these observations suggests that not all of the galactic DM can be in the form of compact DSs; current limits indicate that at most 1–10\% of the DM could be composed of such objects, with the exact fraction depending on their mass and mass spectrum~\cite{Green:2020jor}.
Note also that there is a possibility of asymmetric DM coexisting with baryonic matter in some form of an admixed star~\cite{Leung:2011zz,Leung:2012vea,Kain:2021hpk,Miao:2022rqj,Ivanytskyi:2019wxd,Karkevandi:2021ygv,Leung:2022wcf}. 

Beyond their gravitational signatures, DSs could manifest through other astrophysical signals. For instance, if DM particles kinetically mix with photons, the stars could convert dark photons into observable ordinary photons, potentially producing a distinct luminosity spectrum if photons acquire an effective mass inside DSs~\cite{Maselli:2019ubs}. Alternatively, there could be signatures arising from a slight violation of the dark charge that stabilizes the asymmetric DM. If this quantum number is broken by one or two units, it would allow for the slow decay or annihilation of DM particles into Standard Model particles. Within the dense interior of a DS, even a small breaking of this symmetry, potentially down to the Planck scale, could lead to a significant rate of energy injection~\cite{Betancourt_Kamenetskaia_2023}. This could, in turn, produce detectable signals in current gamma-ray or neutrino telescopes. Furthermore, the energy released by such annihilations could trigger radial oscillations in the DS, leading to a modulation of its luminosity~\cite{Kouvaris:2025ijy}.

Since the detection of the first GW event, GW150914 from a binary black hole (BBH) \cite{LIGOScientific:2016aoc}, GWs from the merger of compact object binaries offer us an important tool to probe the nature of compact objects and DM physics. For instance, the merger of a binary neutron star event, GW170817, provides measurements of neutron star radii and information on its equation of state \cite{LIGOScientific:2018cki}. In addition, the DM population surrounding a compact object could produce unique GW signals~\cite{Ding:2025hqf}. A natural probe to study the nature of DSs is GWs from the merger of a binary of DSs (BDS). A potential detection of their GW signals in future GW interferometers such as LISA \cite{LISA:2017pwj} and DECIGO \cite{Kawamura:2006up}, would allow the extraction of rich information on DSs, like their mass and compactness, and perhaps their corresponding DM particle parameters.

To pursue the BDS merger, we focus on the center of DM halos, where DM spikes probably exists around supermassive black hole (SMBHs). The presence of a strong gravitational attraction around SMBHs could lead DM to adiabatically form a  spike, with a DM density several orders of magnitude larger than that of ordinary DM halos without an SMBH \cite{Gondolo:1999ef}. Although the existence of a DM spike is not conclusively confirmed yet, several studies provide hints of a DM spike surrounding SMBHs~\cite{2024ApJ...962L..40C, Deb:2025raq}. The DS number density will follow the DM distribution and therefore if such a DM spike occurs,  a large number of DSs is expected to exist around an SMBH. In turn this  leads to a significant BDS merger rate that can potentially be detected with future GW interferometers. 

In this work, we study the merger rate of BDS within DM spikes and its detectability in the future space-based GW detectors LISA and DECIGO. The DS parameters, such as DS mass and compactness, depend on the intrinsic DM parameters like the masses of the DM particle and the mediator of the DM self-interactions as well as their respective coupling constant. In addition, the same DM parameters determine the  formation history and the abundance of DSs as a function of the cosmological redshift. Incorporating the formation of DM spikes around SMBHs, we find novel features in the redshift evolution of BDS mergers within these spikes that can be probed by GW observations. We use three  benchmark parameter points to demonstrate this redshift evolution of the BDS merger rate and corresponding DS physics that can be detected in future GW experiments. Two among them could produce a merger rate larger than Population I and II BH mergers.

This paper is organized as follows: in Sec.~\ref{sec:DM_spike}, we introduce the DM spike density profile and populations. In Sec.~\ref{sec:formation}, we discuss the DS formation mechanism. In Sec.~\ref{sec:structure}, we discuss the DS structure and merger cross section. In Sec.~\ref{sec:GW_signals}, we calculate the DS merger rate in DM spikes and its detectability in LISA and DECIGO. In Sec.~\ref{sec:conclusion}, we summarize the results and present our conclusions. Throughout this paper we will assume a spatially flat universe described by the $\Lambda$CDM model with $H_0=100h\, \mathrm{km}\,\mathrm{s}^{-1}\,\mathrm{Mpc}^{-1}$, $h = 0.6736$, $\Omega_{\rm m,0}=0.315$, $\Omega_{\Lambda,0}=0.685$ where the DM corresponds to $\Omega_{\rm DM,0}=0.264$~\cite{Planck:2018vyg}. We will also work in natural units with $c=\hbar=k_B=1$, while keeping $G$ explicit.

\section{Dark matter spike}\label{sec:DM_spike}

The DM spike is a hypothetical DM distribution in the vicinity of SMBHs. It is formed from the adiabatic growth of DM halo density in the gravitational potential of SMBHs and contributes a significant DM density around SMBHs \cite{Gondolo:1999ef}. The existence of a large density DM spike around SMBHs would cause various observational signals, such as DM annihilation \cite{Gondolo:1999ef, Balaji:2023hmy}, primordial black hole (PBH) mergers \cite{Nishikawa:2017chy, Ding:2024mro}, and BH shadows \cite{Chen:2024nua}.

The existence of DM spikes is still under debate \cite{Ullio:2001fb}. Several studies suggest their potential presence around the SMBH OJ\,287 \cite{2024ApJ...962L..40C, Deb:2025raq} and the stellar mass BHs A0620-00 and XTE J1118+480~\cite{Chan:2022gqd}, based on analyzes of the binary orbital decay rate, although alternative explanations remain possible~\cite{Ding:2025nxe}. Since a DM spike can enhance the merger rate of compact binaries such as those of PBHs~\cite{Ding:2024mro} and BDSs that we study here, future GW observations would be able to definitively confirm the existence of the spike in the vicinity of SMBHs. In what follows, we discuss the density profile of a DM spike and its population in the Universe, and use it in the BDS merger rate calculation for the estimate of the GW signal.

\subsection{Dark matter spike density profile}
Due to self-gravitational attraction, DM forms a halo structure in the late-time structure formation and favors a cuspy density distribution in N-body simulations, which is well described by a Navarro-Frenk-White (NFW) profile \cite{Navarro:1995iw, Navarro:1996gj} 
\begin{equation}\label{eq:NFW_profile}
    \rho_{\rm DM}(r) = \frac{\rho_0}{(r/r_0)(1+r/r_0)^2}~,
\end{equation}
where $\rho_0$ is the characteristic density and $r_0$ is a scale length. For the Milky Way, $\rho_0 = 6.6 \times 10^6 \, {\rm M_\odot \, kpc^{-3}}$ and $r_0 = 19.1 \, {\rm kpc}$. In general, the DM density in the galactic center follows a power-law distribution $\rho_{\rm DM}(r) \simeq \rho_0 (r_0/r)^\gamma$, and the power index $\gamma$ can be between $0$ and $2$,  with $\gamma = 1$ corresponding to the NFW profile. The high DM density in the galactic center increases the number density of DSs (which is expected to follow the DM profile) and the corresponding merger rate, offering an ideal laboratory to study DS mergers.

In the galactic center, however, there is an SMBH, whose strong gravitational attraction might modify the surrounding DM distribution. The DM density profile around an SMBH has been studied in detail in Ref.~\cite{Gondolo:1999ef}, suggesting that the DM distribution follows a spike-like profile of the form
\begin{equation}\label{eq:spike_profile}
    \rho_{\rm sp} = \rho_R \left(1 - \frac{4 R_s}{r}\right)^3 \left(\frac{r_{\rm sp}}{r}\right)^{\gamma_{\rm sp}}~,
\end{equation}
where $\rho_R$ is the density at the radius of the DM spike $r_{\rm sp}$ that equals $\rho_0 (r_0/r_{\rm sp})^\gamma$, $R_s = 2 G M_{\rm BH}$ is the Schwarzschild radius of the SMBH with a mass $M_{\rm BH}$ at the galactic center, $G$ is the gravitational constant and $\gamma_{\rm sp} = (9 - 2 \gamma)/(4 - \gamma)$ is the power index of the DM spike. The properties of the DM spike depend on the SMBH mass and the power index of the inner DM halo. The DM spike radius can be calculated as
\begin{equation}\label{eq:rsp}
    r_{\rm sp}(\gamma, M_{\rm BH}) = \alpha_\gamma r_0 \left(\frac{M_{\rm BH}}{\rho_0 r_0^3}\right)^{1/(3-\gamma)}~,
\end{equation}
where $\alpha_\gamma$ is a normalization factor for a given value of $\gamma$ (for its numerical value, see Ref.~\cite{Gondolo:1999ef}). Further analysis in the context of general relativity suggests an inner extended DM spike solution that can form a larger spike density around the Schwarzschild radius $R_s = 2 G M_{\rm BH}$ \cite{Sadeghian:2013laa, Speeney:2022ryg}. However, this relativistic DM spike solution doesn't significantly modify the DS merger rate and therefore in this work we will use the  non-relativistic solution we mentioned above. 

\subsection{Dark matter spike population}
The DM spike density profile describes the DM distribution in the vicinity of SMBHs, and its properties would determine the DS number density and corresponding DS merger rate. In order to obtain the overall DS merger rate, we should consider the contribution of all the relevant DM spikes. Therefore we need to estimate the distribution of DM spikes in the Universe.

Since the properties of DM spikes depend on the mass of their host SMBH, we need an estimate of the SMBH mass function. The SMBH mass is related to parameters of its hosted DM halo via an $M_{\rm BH} - \sigma$ relation, with $\sigma$ being the velocity dispersion of DM halos. We can use the DM halo mass function and the $M_{\rm BH} - \sigma$ relation to obtain the SMBH mass function.
A general $M_{\rm BH}-\sigma$ relation as a function of redshift can be described as \cite{Robertson:2005fh}
\begin{equation}\label{eq:M-sigma}
    \log_{10} \left(\frac{M_{\rm BH}}{M_\odot}\right) = a + b \log_{10} \left(\frac{\sigma}{200 \, {\rm km \, s^{-1}}}\right) - \xi \log_{10} (1+z)~,
\end{equation}
where parameters $a = 8.12 \pm0.08$, $b = 4.24 \pm 0.41$ and $\xi = 0.186$ are empirically determined in Refs.~\cite{Robertson:2005fh, 2009ApJ...698..198G}. Notice that the redshift evolution of the $M_{\rm BH} - \sigma$ relation in Eq.~\eqref{eq:M-sigma} is valid up to redshift $z \sim 6$ in Ref.~\cite{Robertson:2005fh}. To estimate the velocity dispersion $\sigma$ from the NFW profile, we first calculate the velocity dispersion $\sigma$ that  corresponds to the maximal circular velocity at radius $r_m = c_m r_0$ with $c_m = 2.16$ as
\begin{equation}\label{eq:velocity_dispersion}
    \sigma^2 = \frac{GM(c_m r_0)}{c_m r_0} = \frac{4 \pi G \rho_0 r_0^2 g(c_m)}{c_m}~,
\end{equation}
where we used that the enclosed mass $M(r)$ up to a radius $r$ in the NFW profile is $M(r) = 4 \pi \rho_0 r_0^3 g(r/r_0)$ with $g(x) = \log(1+x) - x/(1+x)$. For given SMBH mass and redshift, Eqs.~\eqref{eq:M-sigma} and~\eqref{eq:velocity_dispersion}, relate $\rho_0$ to $r_0$. In the standard spherical collapse model, an overdensity decouples from the Hubble expansion at some turn-around time $t_{\rm ta}$ and it is expected that the collapsing mass virializes by $2t_{\rm ta}$ within the virial radius $r_{\rm vir}$. At that point the density contrast is about $\sim 200$. We can now find an extra relation between 
 $\rho_0$ and $r_0$, by demanding that the virial mass of the DM halo $M_{\rm vir}$ becomes equal to the halo mass which follows an NFW profile truncated at $r_{\rm vir}$ 
\begin{equation}\label{eq:vir_mass}
    M_{\rm vir} 
    = 200 \rho_{\rm crit} \left(\frac{4 \pi (c(M_{\rm vir}) r_0)^3}{3}\right)
    =4\pi \rho_0 r_0^3 g(c(M_{\rm vir}))~, 
\end{equation}
where $\rho_{\rm crit}$ is the critical energy density of the Universe at a given redshift, and $c(M_{\rm vir}) \equiv r_{\rm vir}/r_0$ is the concentration parameter. The numerical solution of $c(M_{\rm vir})$ can be calculated from Ref.~\cite{2012MNRAS.423.3018P}.
This condition relates $\rho_0$ and $r_0$ and along with the condition from 
  Eqs.~(\ref{eq:M-sigma}--\ref{eq:velocity_dispersion}), we have two equations that relate the two quantities. Consequently we can now determine the value of both $\rho_0$ and $r_0$. Once the parameters are known, we can use them in Eq.~\eqref{eq:vir_mass} to estimate the virial mass of the DM halo. Based on the above discussion, the SMBH mass is related to its host DM halo at a given redshift, and this relation connects the SMBH mass function with the DM halo mass function.

To obtain the SMBH mass function, we first calculate the DM halo mass function $dn/dM_{\rm vir}$, as estimated in Ref.~\cite{2012MNRAS.423.3018P}. It reads
\begin{equation}
    \frac{dn}{dM_{\rm vir}} = f(\sigma_M) \frac{\rho_m}{M_{\rm vir}} \frac{d \log (\sigma_M^{-1})}{d M_{\rm vir}}~,
\end{equation}
where $\rho_m$ is the redshift-dependent cosmological matter density with a present value of $39.7\, M_\odot\,{\rm kpc^{-3}}$ and $\sigma_M$ is the linear root-mean-square fluctuation of the density field on the scale $M_{\rm vir}$ that can be semi-analytically calculated from Ref.~\cite{2011ApJ...740..102K}. The $f(\sigma_M)$ function represents the collapsing overdense regions that is well described by
\begin{equation}
    f(\sigma_M) = A \left(1 + \left(\frac{\sigma_M}{b}\right)^{-a}\right) \exp \left(- \frac{c}{\sigma_M^2}\right)~,
\end{equation}
with parameters $A = 0.213$, $a = 1.8$, $b=1.85$ and $c = 1.57$ corresponding to a spherical collapse case~\cite{Tinker:2008ff}. We can now use the relation between the SMBH mass and the DM halo mass to estimate the SMBH mass function as
\begin{equation}
    \frac{dn}{dM_{\rm BH}} = \frac{dn}{dM_{\rm vir}} \frac{dM_{\rm vir}}{dM_{\rm BH}} = f(\sigma_M) \frac{\rho_m}{M_{\rm vir}} \frac{d \log (\sigma_M^{-1})}{d M_{\rm vir}} \frac{dM_{\rm vir}}{dM_{\rm BH}}~.
\end{equation}
Here, the differential relation $d M_{\rm vir}/dM_{\rm BH}$ can be obtained from the above discussed relation between SMBH mass and DM halo mass. Since the DM spike properties depend on the SMBH mass, this SMBH mass function gives the statistical information on the DM spike population.

\section{Dark star formation}\label{sec:formation}
DM halos provide the ideal setting for BDS mergers, but the properties and cosmological abundance of the DSs themselves are determined by the underlying dark sector microphysics, which we now describe. Following the framework introduced in Ref.~\cite{Chang:2018bgx}, we consider a dark sector composed of two particle species: a massive dark electron of mass $m_{e_D}$, which constitutes a fraction $f_{e_D}$ of the total DM abundance, and a light dark photon of mass $m_{\gamma_D}$ that mediates long-range interactions among dark electrons. The interaction strength is parameterized by the dark fine-structure constant $\alpha_D = g_D^2/(4\pi)$, where $g_D$ denotes the associated gauge coupling. In this setup, dissipative dynamics in the dark sector enable efficient cooling within DM halos, leading to the formation of compact DSs, as detailed in~\cite{Chang:2018bgx}. The resulting cosmological abundance of such DSs has been computed in Ref.~\cite{BetancourtKamenetskaia:2025irn}. In this section, we summarize the relevant model parameters, as well as the physical mechanisms and characteristic timescales governing their formation.

\subsection{DM model parameters}
This class of models is subject to observational constraints from astrophysical and cosmological probes. The most relevant bounds arise from observations of galaxy cluster mergers (\textit{e.g.}, the Bullet Cluster) and the shapes of DM halos, which limit the self-scattering cross-section~\cite{Tulin:2017ara,Rocha:2012jg}. For the case where all DM consists of dark electrons ($f_{e_D}=1$), the momentum-transfer cross-section must satisfy
\begin{equation}
    \frac{\sigma_M}{m_{e_D}}\approx4\pi\frac{\alpha_D^2 m_{e_D}}{m_{\gamma_D}^4}\lesssim 1~\mathrm{cm}^2/\mathrm{g}.
    \label{eq:Moller-cross-section}
\end{equation}

DSs are expected to form through the contraction and subsequent fragmentation of primordial overdensities composed of dark electrons. To model the formation history of these objects, we consider a spherical overdensity region of cold DM (CDM), that is a proto-halo that undergoes gravitational collapse simultaneously with the embedded dark electron component. We write the total CDM mass contained within this proto-halo as $M_{\text{halo}}$, which yields a mass in the form of dark electrons $M_{e_D} = f_{e_D} M_{\text{halo}}$. The evolution of the proto-halo densities is described within the framework of the spherical collapse model~\cite{book_Mo_vdBosch_White}. In this picture, overdense regions gradually decouple from the Hubble expansion and eventually reach a maximum radius, the so-called turn-around point, at a characteristic redshift $z_{\text{ta}}$. At this moment, the density of the CDM component is given by
\begin{equation}
\rho_{\text{DM}}(z_{\text{ta}}) = \frac{9\pi^2}{16} \bar{\rho}_{\text{DM}}(z_{\text{ta}}), \label{eq:turnaround}
\end{equation}
where $\bar{\rho}_{\text{DM}}(z)$ denotes the background DM density in the Universe at redshift $z$. Subsequently, the dark electron component has a density of $\rho_{e_D}(z_{\text{ta}}) = f_{e_D} \rho_{\text{DM}}(z_{\text{ta}})$, with a number density $n_{e_D}(z_{\text{ta}}) = \rho_{e_D}(z_{\text{ta}})/m_{e_D}$ at turn-around. We further assume that at turn-around the dark electron gas possesses an initial temperature $T_{e_D}(z_{\text{ta}})$, which we treat as a free parameter (though our conclusions are not strongly sensitive to its precise value). For concreteness, we set $T_{e_D}(z_{\text{ta}})=5\times10^{-3} T_{\rm CMB}(z_{\rm ta})$, where $T_{\rm CMB}=2.7255\, \mathrm{K}(1+z)$ is the temperature of the cosmic microwave background (CMB).

For the dark electron gas to undergo further collapse beyond that of the surrounding CDM, it must become effectively self-interacting. This requires two simultaneous conditions to be satisfied: (i) the mean free path of dark electrons must be smaller than the characteristic size of the collapsing CDM overdensity, ensuring that dark electrons interact frequently enough to thermalize and lose energy; and (ii) the mean free path of dark photons must remain larger than this size, so that dark photons can escape freely and carry away energy from the system, rather than staying within the perturbation and forming a coupled fluid in equilibrium with the dark electrons. The first condition translates into an upper limit on the dark photon mass \citep{Chang:2018bgx}:
\begin{equation}\label{eq:m_gamma_D_chi_P}
    m_{\gamma_D}\lesssim 3.8\times10^8~\mathrm{eV}\left(\frac{1+z}{1+3400}\right)^{\frac12}\left(\frac{\alpha_D}{0.1}\right)^{\frac12}\left(\frac{f_{e_D}}{1}\right)^{\frac14}\left(\frac{m_{e_D}}{1~\rm GeV}\right)^{\frac14}\left(\frac{M_{\rm halo}}{10^{14}M_\odot}\right)^{\frac{1}{12}}.
\end{equation} 
This constraint implies that not all of the DM can consist of dark electrons. In particular, combining Eq.~(\ref{eq:m_gamma_D_chi_P}) with the M{\o}ller scattering cross section in Eq.~(\ref{eq:Moller-cross-section}) would require halo masses $M_{\rm halo}\gtrsim 10^{17}\,M_\odot$ at $z\lesssim 30$ if $f_{e_D}=1$, in conflict with the observed upper limit on halo masses of ${\cal O}(10^{15}\,M_\odot)$ at the present epoch. The second condition translates into a lower limit on the dark electron mass:
\begin{equation}
m_{e_D}\gtrsim3.7~\mathrm{MeV}\left(\frac{1+z}{1+30}\right)^{\frac23}\left(\frac{\alpha_D}{0.1}\right)^{\frac23}\left(\frac{f_{e_D}}{1}\right)^{\frac13}\left(\frac{M_{\rm halo}}{10^{10}M_\odot}\right)^{\frac19}.
\end{equation}
If $m_{e_D}$ were smaller than this bound, Compton scattering would tightly couple dark electrons and dark photons, causing them to behave as a single fluid. In that regime, radiative cooling would be inefficient, preventing energy loss and halting the collapse and fragmentation necessary for DS formation.

\subsection{Dark star formation within a DM halo}
To model the evolution with time of the dark electron clump we use the first law of thermodynamics, which dictates that
\begin{equation}
\frac{dE}{dt}=-P_{e_D}\frac{dV}{dt}-\Lambda V,
\label{eq:dEdt}
\end{equation}
where $E$ is thermal energy of the dark electron gas enclosed in the volume $V$, where the pressure of the gas of dark electrons with photon-induced self-interactions is~\cite{2015PhRvD..92f3526K}
\begin{equation}\label{eq:self_int_pressure}
    P_{e_D}=n_{e_D}T_{e_D}+2\pi\alpha_Dn_{e_D}^2/m_{\gamma_D}^2.
\end{equation}
Finally, $\Lambda$ is the rate of energy loss per unit volume. For dark electron bremsstrahlung, the total energy emission rate per unit volume in the non-relativistic regime is given by \citep{1975ZNatA..30.1546H,Chang:2018bgx}
\begin{equation}\label{eq:dark_bremss}
    \Lambda_{\rm \gamma_D}=\frac{32\alpha_{D}^3 n_{e_D}^2 T_{e_D}}{\sqrt{\pi}m_{e_D}^2} \sqrt{\frac{T_{ e_D}}{m_{e_D}}}\mathrm{e}^{-\frac{m_{\gamma_D}}{T_{e_D}}},
\end{equation}
where the exponential factor accounts for the suppression due to the nonzero mass of the dark photon when $m_{\gamma_D}\geq T_{e_D}$. The rate of energy loss per unit mass is 
\begin{equation}\Lambda=\Lambda_{\gamma_D}\, \mathrm{exp}\left[-R \sigma_{\rm C}n_{e_D}\right],
\label{lambda}
\end{equation}
where the extra exponential suppression incorporates the drop in the rate due to the random walk of dark photons that scatter via a Compton-like interaction in case their mean free path is smaller than the size of the clump. Here, $R$ is the radius of the clump and $\sigma_{\rm C}=\frac{8\pi}{3}\frac{\alpha_D^2}{m_{e_D}^2}$ is the Compton cross-section for dark photons scattering off dark electrons.

Using that the energy of the homogeneous dark electron gas enclosed in volume $V$ is $E=(3/2) n_{e_D} V T_{e_D}$ and that the volume is related to the dark electron number density through $V=M_{e_D}/(n_{e_D}m_{e_D})$, Eq.~(\ref{eq:dEdt}) can be recast as:
\begin{equation}
    \frac{d\ln{T_{e_D}}}{d\ln{n_{e_D}}}=\frac{2}{3}\frac{P_{e_D}}{n_{e_D}T_{e_D}}-2\left[\frac{\Lambda}{3n_{e_D}T_{e_D}}\left(\frac{d\ln n_{e_D}}{dt}\right)^{-1}\right].
    \label{lnT}
\end{equation}

The collapse of a dark electron clump from its initial turn-around state to compact DSs proceeds through three distinct evolutionary stages, each characterized by a different balance between gravitational contraction, pressure support, and radiative cooling. The transition between these stages is determined by the relative efficiency of cooling via dark photon bremsstrahlung compared to the dynamical timescales of the system. In the following, we will summarize them and contextualize the physical process in terms of the quantities needed to solve Eq.~\eqref{lnT}.

\paragraph{1st stage of collapse: adiabatic free-fall.}
In the standard simplified picture of structure formation, primordial overdensities will grow during the matter domination phase of the Universe and become nonlinear eventually decoupling from the Hubble expansion. After the expansion is halted, the overdensity starts collapsing  leading eventually to the formation of a virialized halo.
In our setup the main CDM component collapses carrying along both baryons and the minor strongly interacting dark electron component. The CDM component is assumed to be non-interacting other than gravity. The dark electron component, however, develops ideal gas pressure. Since at the beginning of the evolution, the halo is still  dilute, the interactions among dark electrons are rare and the energy loss via the bremsstrahlung of dark photons is practically negligible. Therefore, the collapse can be considered as adiabatic with pressure $P_{e_D}\approx n_{e_D}T_{e_D}$, and the evolution of the temperature of the dark electron gas is simply related to the dark electron density through:
\begin{equation}\label{eq:T_ff}
    T_{e_D}(z)=T_{\rm e_D}(z_{\rm ta})\left(\frac{n_{e_D}(z)}{n_{e_D}(z_{ta})}\right)^{\frac23}.
\end{equation}
This is the first stage of free-fall and corresponds to setting $\Lambda\sim0$ in Eq.~\eqref{lnT}.

\paragraph{2nd stage of collapse: nearly virialized contraction (nvc).} The initial adiabatic collapse stops at the temperature and density $(T^J_{e_D},n^J_{e_D})$, when the mass of the dark electron component in the collapsing halo, $M_{e_D}$, becomes equal to its Jeans mass, given by~\cite{Chang:2018bgx}
\begin{equation}\label{eq:const_jeans_mass}
     m_J=\frac{\pi}{6}\left(\frac{\pi}{G}\right)^{3/2}c_s^3(m_{e_D}n_{e_D})^{-\frac{1}{2}},
\end{equation}
where $c_s^2=\partial P_{e_D}/\partial \rho_{e_D}$ is the square of the speed of sound. At this stage of collapse, self-interactions are subdominant so we may approximate $c_s\approx \sqrt{T_{e_D}/m_{e_D}}$ (with $T_{e_D}$ given by Eq.~(\ref{eq:T_ff})). Once the Jeans mass becomes equal to the total mass of the dark electron component, the latter evolves at first through a series of quasi-virialized states. The  dark bremsstrahlung radiation rate is suppressed at the beginning due to the fact that the collapsing halo is still dilute. Nevertheless, the small energy loss of the system via dark bremsstrahlung, forces the dark electron cloud to re-virialize rapidly and contract again, thus effectively forcing the dark electron clump to move along a line of constant Jeans mass in the density-temperature phase space. We will name this phase as the ``nearly virialized contraction'' (nvc) stage.  Using Eqs.~(\ref{eq:const_jeans_mass}) and (\ref{eq:T_ff}), the equality of the mass of the dark electron gas to the Jeans mass implies that the temperature $T_{e_D}$ of the dark electrons at the transition must be related to their density through:
\begin{align}
\label{eq:T_nvc}
    T_{e_D}=\left(\frac{6}{\pi}\right)^{\frac23}\left(\frac{G}{\pi}\right)m_{e_D}^{\frac43}f_{e_D}^{\frac23}M_{\rm halo}^{\frac23}{n_{e_D}}^{\frac13}.
\end{align}
In particular, this relation defines the location of the point $(T^J_{e_D},n^J_{e_D})$. In this phase 
$\frac{d\ln{T_{e_D}}}{d\ln{n_{e_D}}}=\frac13$, therefore using Eq.~(\ref{lnT}) and that the change in the internal energy of the gas is dominated by the energy loss via bremsstrahlung, one obtains that the dark electron density changes with the time as $\left(\frac{d\ln n_{e_D}}{dt}\right)^{-1}=\frac{n_{e_D}T_{e_D}}{2\Lambda}$.

\paragraph{3rd stage of collapse: Fragmentation.}
Lastly, as the temperature and density of the clump increase, so does the bremsstrahlung loss rate. At large enough densities, the energy loss timescale becomes comparable to the so-called free-fall timescale, defined as  $t_{\rm ff}=\left(16\pi G \rho_{e_D}\right)^{-1/2}$~\cite{Penston:1969yy}. This marks the onset of the final phase named ``fragmentation'', which occurs at the temperature and density  $(T^{\rm frag}_{e_D},n^{\rm frag}_{e_D})$, determined by
\begin{equation}
\label{eq:free-fall_time}
   \frac{2\Lambda}{3n^{\rm frag}_{e_D}T^{\rm frag}_{e_D}}= \left(\frac{d\ln n_{e_D}}{dt}\right)\Big|_{n_{e_D}=n_{e_D}^{\rm frag}} = \left(16\pi G m_{e_D}n_{e_D}^{\rm frag}\right)^{1/2}.
\end{equation}
In this stage, the temperature of the dark electron gas depends on the density as $T_{e_D}\propto n_{e_D}^{-4/3}$, as follows from  Eqs.~\eqref{lnT} and \eqref{eq:free-fall_time}, and the energy is rapidly evacuated as the clump continues collapsing, ultimately fragmenting to several self-gravitating chunks of matter with mass equal to the Jeans mass. This happens because during this phase, the temperature drops and this leads to a reduction of the Jeans mass. Therefore the initial collapsing dark electron cloud fragments into several self-gravitating chunks; each of them with a mass equal to the Jeans mass at the given time.

Eventually, the fragmentation stops due to one of the following reasons: i) each fragmented collapsing piece becomes degenerate and develops Fermi pressure, or ii) dark electron repulsive self-interactions forbid further collapse, or iii) because the density of dark electrons has increased to the extent that the mean free path of dark photons has become smaller than the size of the collapsing fragment. In this last case the latter becomes optically thick. This occurs when the exponential term of Eq.~\eqref{lambda} starts dominating and dark photons are reabsorbed before they can escape from the clump:
\begin{align}
R\sigma_c n_{e_D}>1.
\end{align}
The energy is not evacuated from the bulk, the temperature stops dropping and the formed objects continue cooling only via emission from the surface. This marks the formation of the minimal fragments, that is, the DSs.

\begin{figure}[t!]
	\centering
	\includegraphics[width=.49\textwidth]{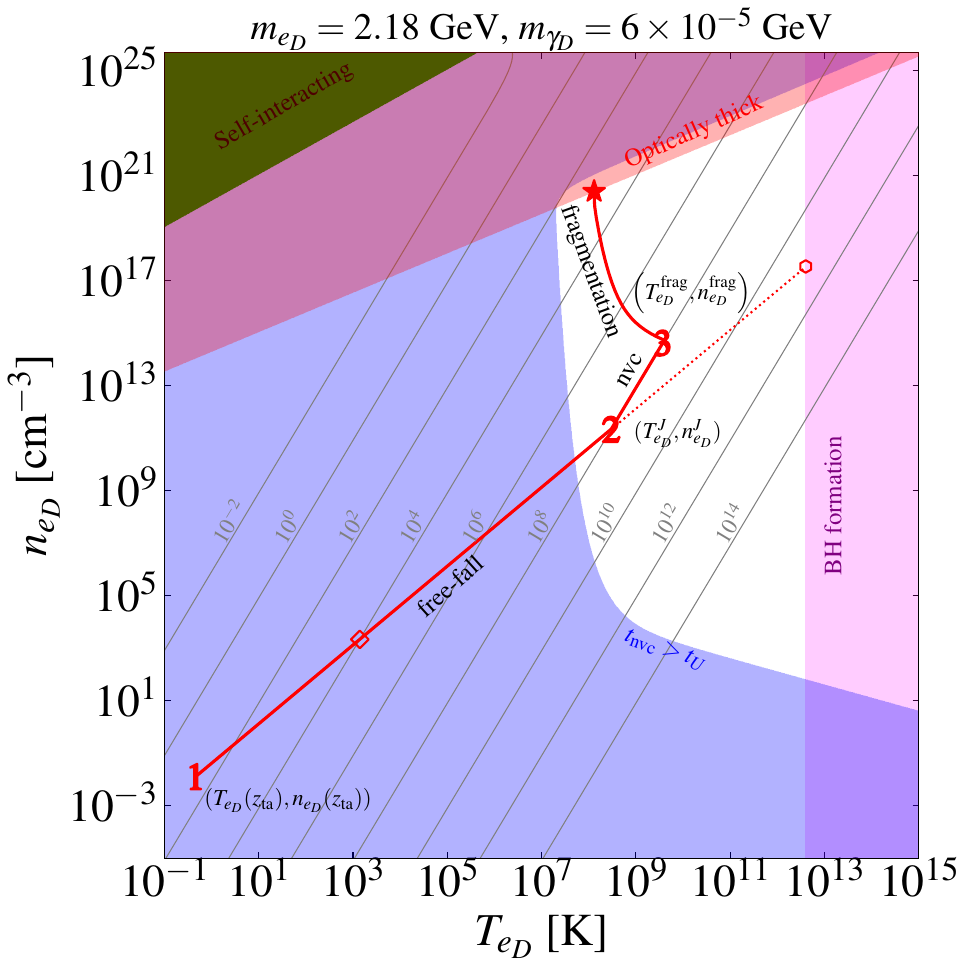}
    \includegraphics[width=.49\textwidth]{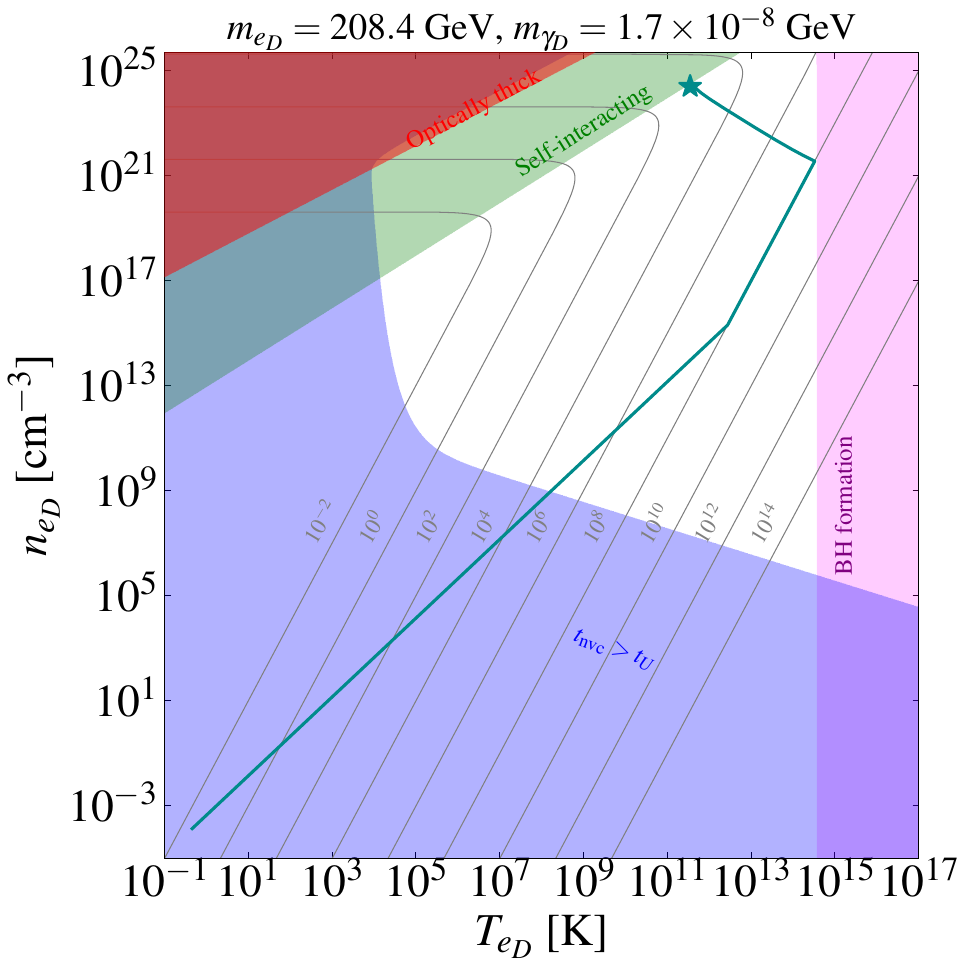}
	\caption{\textit{Left panel}: Number density vs temperature of a dark electron clump in  
    a halo with $M_{\rm halo}=10^{5}M_\odot$ (red rhombus), $M_{\rm halo}=10^{9}M_\odot$ (red solid line) and $10^{13}M_\odot$  (red dotted line), assuming that 10\% of the total mass of the halo is in the form of dark electrons. We take the parameter values: $m_{e_D}=2.18~\rm GeV$, $m_{\gamma_D}=60~\rm keV$ and $\alpha_D=0.1$. Solid gray curves: Contours of constant Jeans mass. Red-shaded region: Optical thickness region $R\sigma_c n_{e_D}>1$. Green-shaded region: Region where self-interaction pressure overcomes kinetic pressure $n_{e_D}T_{e_D}<2\pi\alpha_Dn_{e_D}^2/m_{\gamma_D}^2$. Purple-shaded region: BH formation region. Blue-shaded region: Region where bremsstrahlung cooling is inefficient and DSs would not have been formed by today. \textit{Right panel}: Same as left, but for $m_{e_D}=208.4~\rm GeV$, $m_{\gamma_D}=17~\rm eV$ and $\alpha_D=0.1$.}
	\label{fig:Formation_phase_space}
\end{figure}

In Fig.~\ref{fig:Formation_phase_space} we sketch the evolution of the dark electron clump number density and temperature for two choices of parameters. In the left panel, we consider $m_{e_D}=2.18~\rm GeV$, $m_{\gamma_D}=60~\rm keV$, $\alpha_D=0.1$ and $f_{e_D}=10\%$, while in the right panel, we take $m_{e_D}=208.4~\rm GeV$,  $m_{\gamma_D}=17~\rm eV$, $\alpha_D=0.1$ and $f_{e_D}=10\%$. Both of these choices yield a DS mass of $1000M_\odot$. The solid gray curves correspond to contours of constant Jeans mass in units of $M_\odot$.

In the left panel, we present two different choices of the halo mass: for the solid red curve we consider $ M_{\rm halo}=10^{9} M_\odot$, while for the dotted red line we set $ M_{\rm halo}=10^{13} M_\odot$. We fix the initial condition in both cases with the turnaround temperature and density represented by the red ``1''. For our analysis we adopt $T_{e_D}(z_{\rm ta})=5\times10^{-3}T_{\rm CMB}(z_{ta})$ (while $T_{\rm CMB}(z_{\rm ta})\simeq 2.73 \,{\rm K}\,(1+z_{ta})$), although our conclusions do not depend strongly on the specific value. We focus first on the red solid curve, which represents a scenario of possible DS formation. Starting from the initial condition, the clump goes through the adiabatic free-fall stage until the Jeans mass becomes equal to the total dark electron mass $m_J=f_{e_D}M_{\rm halo}=10^8M_\odot$. This marks the transition to the nvc stage and is indicated by the red ``2''. We have chosen the turnaround redshift such that the red point is reached at $z=21$ (giving $z_{\rm ta}\sim33$). The nvc stage continues until the transition to the fragmentation phase, which is denoted by the red ``3''. Finally, fragmentation stops when the dark electron clump becomes optically thick and is presented as the intersection of the solid red trajectory with the red-colored region ``optically thick'', denoted by the red star-shaped point in the figure. 

It is important to emphasize that not all DM halos will successfully produce DSs. The fate of a collapsing dark electron clump depends sensitively on the halo mass, with two regimes where DS formation is suppressed: very massive halos lead to BH formation, while very light halos experience cooling timescales that exceed the age of the Universe. For sufficiently massive halos, the contraction phase can produce such extreme densities that the clump collapses directly into a BH before fragmentation can occur. This happens when the radius of a clump becomes smaller than its Schwarzschild radius $R_s=2G m_J$. Using the Jeans mass expression from Eq.~(\ref{eq:const_jeans_mass}) and the condition $R < R_s$, we obtain a critical temperature threshold $T_{e_D}\gtrsim T^{\rm sch}_{e_D}$ with 
\begin{equation}
 T^{\rm sch}_{e_D}= 6\times10^{12}~\mathrm{K}\left(\frac{m_{e_D}}{1~\rm GeV}\right),
\end{equation}
which is indicated in the figure as a purple band. As apparent from the plot, a dark electron clump will end up as a BH when $T^{\rm frag}_{e_D}\gtrsim T^{\rm sch}_{e_D}$. The temperature at which fragmentation would otherwise begin, $T_{e_D}^{\text{frag}}$, can be calculated by combining the constant Jeans mass trajectory from Eq.~(\ref{eq:T_nvc}) with the fragmentation condition in Eq.~(\ref{eq:free-fall_time}). Neglecting the exponential suppression factors in the cooling function $\Lambda$, we find: 
\begin{align}
    T_{e_D}^{\rm frag}&\approx \frac{9\pi^2 G}{64}\left[\frac{9\pi^3}{64}\left(\frac{\pi}{6}\right)^{\frac23}\right]^{-\frac34}\alpha_D^{-\frac32}m_{e_D}^{\frac52}f_{e_D}^{\frac12}M_{\rm halo},
\end{align}
When $T_{e_D}^{\text{frag}} \gtrsim T_{e_D}^{\text{sch}}$, the clump collapses to a BH before entering the fragmentation phase. This scenario is illustrated in Fig.~\ref{fig:Formation_phase_space} by the dotted red trajectory corresponding to a halo mass $M_{\text{halo}} = 10^{13}M_\odot$, which crosses into the BH region before fragmentation can occur at the red hexagon point. In general, this behaviour leads to the definition of a maximum halo mass $M^{\rm max}_{\rm halo}$, above which no DSs are expected. We can estimate this mass from Eq.~\eqref{eq:T_nvc} and imposing $R=2Gm_J$ to be
\begin{equation}\label{eq:max_halo_mass}
    M^{\rm max}_{\rm halo}=1.1\times10^{15}M_\odot\left(\frac{\alpha_D}{0.1}\right)^3f_{e_D}^{-1}\left(\frac{m_{e_D}}{1~\rm GeV}\right)^{-3}.
\end{equation}

On the other hand, for very light halos, the contraction and fragmentation proceed too slowly to be completed on cosmological timescales. If the cooling timescale, defined from Eq.~\eqref{lnT} as $t_{\rm cooling}=\left(\Lambda/(3n_{e_D}T_{e_D})\right)^{-1}$, exceeds the Hubble time, the dark electron clump remains trapped in the nearly virialized contraction phase indefinitely, never reaching the densities required for fragmentation. This is shown in Fig.~\ref{fig:Formation_phase_space} by the red rhombus for a halo mass $M_{\text{halo}} = 10^5 M_\odot$. The red rhombus  marks the present-day state of this clump, which lies entirely within the blue-shaded region where bremsstrahlung cooling is too inefficient to drive further collapse. Such halos would contribute to a diffuse component of dark electrons rather than forming compact objects. This suggests the definition of a minimum halo mass $M_{\rm halo}^{\rm m in}$, which we estimate from the implicit Eq.
\begin{equation}\label{eq:M_halo_min_def}
    \left(\frac{d\ln n_{e_D}}{dt}\right)^{-1}\Big|_{\rm nvc}=\frac{n^J_{e_D}(M^{\rm min}_{\rm halo})T^J_{e_D}(M^{\rm min}_{\rm halo})}{2\Lambda(M_{\rm halo}^{\rm min})}=(t_U-t_{\rm nvc}(z)).
\end{equation}
Here $t_U \approx 13.8\,\text{Gyr}$ denotes the present age of the Universe, while $t_{\text{nvc}}(z)$ represents the cosmic time at which the clump first entered the nearly virialized contraction phase at redshift $z$. The difference $t_U - t_{\text{nvc}}(z)$ therefore corresponds to the time elapsed from the onset of the nvc phase until today. The central part of Eq.~\eqref{eq:M_halo_min_def} characterizes the characteristic timescale for DS formation beginning from the nvc stage. Therefore, Eq.~\eqref{eq:M_halo_min_def} is merely the requirement that this formation timescale must not exceed the time available since the nvc phase began, ensuring that DSs can indeed form by the present epoch. 

Finally, the right panel of Fig.~\ref{fig:Formation_phase_space} presents a different phase space trajectory for different DM parameters. The colored regions follow the same convention as in the left panel. The main difference here is the condition for the end of fragmentation. In this case the end of fragmentation is due to the self-interactions of the dark electrons. For small enough dark photon mass, as the density increases and the temperature decreases, the self-interacting
term in the dark electron pressure~\eqref{eq:self_int_pressure} may become dominant over the kinetic term. As a result, the Jeans mass~\eqref{eq:const_jeans_mass} would become independent of the temperature and begin to increase proportional to the density. Therefore, by the Jeans criterion, fragmentation would stop, with the final fragments being once again the DSs. We color the region of phase space where this condition is satisfied in green and denote it as ``self-interacting''. The end of fragmentation is then presented as intersection of the solid cyan trajectory with this region, denoted by the cyan star-shaped point in the figure.

\subsection{Cosmological dark star density} \label{sec:ds_energy_density}
\begin{figure}[t!]
	\centering
    \includegraphics[width=.48\textwidth]{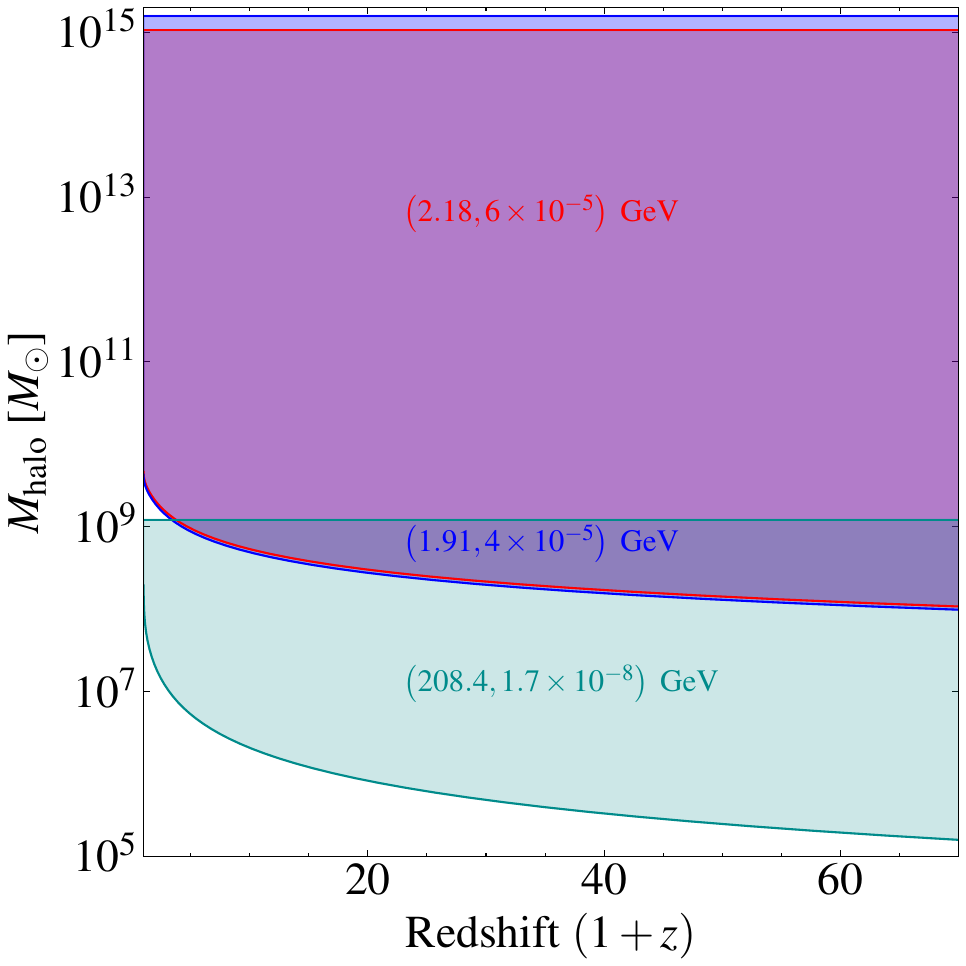}
	\includegraphics[width=.48\textwidth]{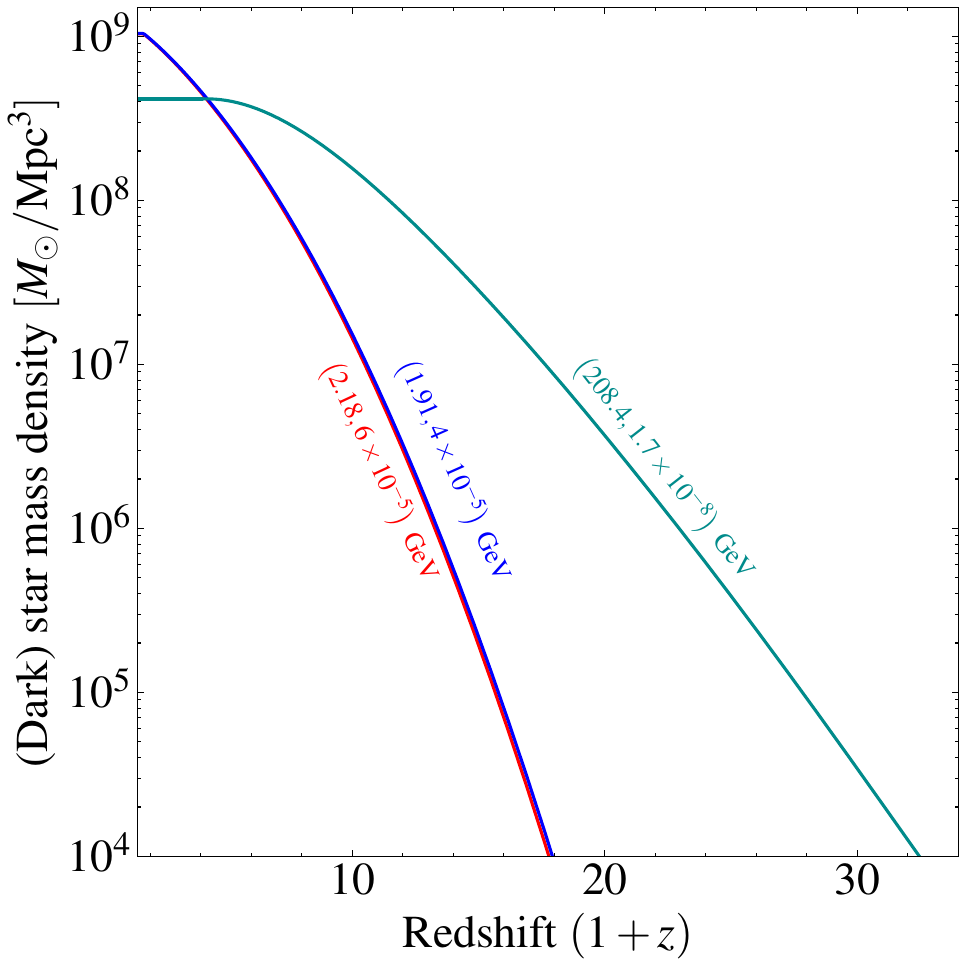}
	\caption{\textit{Left panel:} Range of halo masses leading to DS formation at redshift $z$, for $m_{e_D}=2.18~\rm GeV$ and $m_{\gamma_D}=60~\rm keV$ (red-colored region), $m_{e_D}=1.91~\rm GeV$ and $m_{\gamma_D}=40~\rm keV$ (blue-colored region) and $m_{e_D}=208.4~\rm GeV$ and $m_{\gamma_D}=17~\rm eV$ (cyan-colored region). The rest of parameters are as in Fig.~\ref{fig:Formation_phase_space}. \textit{Right panel:} DS mass density as a function of redshift. The colored curves correspond to the benchmark parameters from the left panel.}
	\label{fig:DSMD}
\end{figure}

We show in the left panel of Fig.~\ref{fig:DSMD} the range of halo masses at a given redshift that lead to the production of DSs, 
for $m_{e_D}=2.18~\rm GeV$ and $m_{\gamma_D}=60~\rm keV$ (red-colored region), $m_{e_D}=1.91~\rm GeV$ and $m_{\gamma_D}=40~\rm keV$ (blue-colored region) and $m_{e_D}=208.4~\rm GeV$ and $m_{\gamma_D}=17~\rm eV$ (cyan-colored region), keeping the remaining parameters as in Fig.~\ref{fig:Formation_phase_space}. All halos with masses in this range will lead to a population of DSs at redshift $z$.

The DS mass density  at redshift $z$  (or alternatively the total mass in form of DSs per comoving volume) can be estimated as:
\begin{align}
    \rho_{\rm DS}(z)=f_{e_D-\rm DS}f_{e_D}\frac{\Omega_{\rm DM,0}}{\Omega_{\rm m,0}}
    \int_{M_{\rm halo}^{\rm min}(z)}^{M_{\rm halo}^{\rm max}(z)}{M\,\frac{dn}{dM}(M)\,dM}.
\end{align}
where $\frac{dn}{dM}(M) dM$ is the comoving number density of halos with masses between $M$ and $M+dM$, $\Omega_{\rm DM,0}/\Omega_{\rm m,0}\simeq 5$ is the ratio of the DM and visible matter density parameters today (and any cosmic epoch), $f_{e_D}$ is the fraction of DM in the form of dark electrons, and $f_{e_D-\rm DS}$ is the fraction of dark electrons in the halo that are in the form of DSs (the rest constitutes a diffuse component). We will model the number density of halos following Press \& Schechter \citep{1974ApJ...187..425P}. In this case, the total mass density of the collapsed halos at redshift $z$ is:
\begin{equation}\label{eq:f_coll_DS}
     \int_{M_{\rm halo}^{\rm min}(z)}^{M_{\rm halo}^{\rm max}(z)}{M\,\frac{dn}{dM}\,dM}=\rho_{\rm m}(z)\left[\mathrm{erfc}\left(\frac{\delta_c(z)}{\sqrt{2}\sigma(M_{\rm halo}^{\rm min}(z))}\right)-\mathrm{erfc}\left(\frac{\delta_c(z)}{\sqrt{2}\sigma(M_{\rm halo}^{\rm max}(z))}\right)\right],
\end{equation}
where $\rho_{m}(z)$ is the total matter density, $\delta_c(z)$ is the linear overdensity at virialization and $\sigma^2(M)$ is the variance of the density field when smoothed on scale $M$, and which are given in \citep{2013A&C.....3...23M}. The right panel of Fig.~\ref{fig:DSMD} shows the DS mass density as a function of the redshift for the considered benchmark values.

Two pertinent observations from this figure: first, while the red and blue benchmarks encompass a broader range of halo masses, their integrated mass density is surpassed by the cyan solution at high redshifts. This is because the halos hosting DS formation in the red and blue scenarios are predominantly massive and, consequently, form later in cosmic history. In contrast, the cyan benchmark favors less massive halos, which are known to collapse and form at earlier epochs. Secondly, in the right panel we observe that the mass density reaches a peak before $z=0$, after which it remains constant (\textit{e.g.}, for the cyan curve). This behavior arises from the presence of the upper mass limit, $M_{\mathrm{halo}}^{\mathrm{max}}$, in the DS model. At low redshifts, more massive halos, which are more likely to exceed $M_{\mathrm{halo}}^{\mathrm{max}}$, become increasingly common, thereby suppressing further growth in the DS mass density.

\section{Dark star structure and merger cross section}\label{sec:structure}
\subsection{Structure}
To estimate the density profile of a DS we use the Tolman–Oppenheimer–Volkoff (TOV) equation, which describes the condition of hydrostatic equilibrium for the pressure  $P(r)$
\begin{equation}\label{eq:DSStellar Structure}
	\begin{aligned}
		&\frac{dP}{dr}=-\frac{G M \rho}{r^2}\left(1+\frac{P}{\rho}\right)\left(1+\frac{4\pi r^3 P}{M }\right)\left(1-\frac{2GM}{r}\right)^{-1}.
	\end{aligned}
\end{equation}
Here, $M(r)$ is the mass enclosed in the sphere of radius $r$, which can be obtained from the mass equation
\begin{eqnarray}
	 \frac{dM}{dr}=4\pi r^2\rho, \label{DSStellarStructureA} 
\end{eqnarray}
where the density distribution is related to the pressure $P(r)$ through the equation of state.
For a gas of dark electrons interacting via a dark photon, the EoS is given in the following parametric form~\citep{2015PhRvD..92f3526K}
\begin{subequations}\label{eq:EOS}
	\begin{align}
		&\rho(x)=m_{e_D}^4\left[\xi(x)+\frac{2\alpha_D}{9\pi^3}\left(\frac{m_{e_D}}{m_{\gamma_D}}\right)^2x^6\right], \\
		& P(x)=m_{e_D}^4\left[\psi(x)+\frac{2\alpha_D}{9\pi^3}\left(\frac{m_{e_D}}{m_{\gamma_D}}\right)^2x^6\right],
	\end{align}
\end{subequations}
where the functions $\xi$ and $\psi$ are given by
\begin{subequations}
	\begin{align}
		& \xi(x)=\frac{1}{8\pi^2}\left[x\sqrt{1+x^2}(1+2x^2)-\ln\left(x+\sqrt{1+x^2}\right)\right], \\
		& \psi(x)=\frac{1}{8\pi^2}\left[x\sqrt{1+x^2}(2x^2/3-1)+\ln\left(x+\sqrt{1+x^2}\right)\right].
	\end{align}
\end{subequations}
Here we have introduced the dimensionless parameter $x=p_F/m_{e_D}$, where $p_F$ is the Fermi momentum of the dark electrons. This parameter measures how relativistic the particles on the Fermi surface are. We solve the TOV and mass equations using the EoS above with the boundary conditions:  i) $M(0)=0$, imposing that the mass distribution is not singular at the origin, ii) choosing a value for $x$ at the core, $x_c \equiv x(r=0)$, which is equivalent to fixing the central pressure $P(0) = P(x_c)$ or the central density $\rho(0) = \rho(x_c)$, and iii) that $P(R_{\rm DS})=0$, namely that the pressure at the boundary of the DS vanishes. The boundary of the DS is given by its radius, $R_{\rm DS}$, which is implicitly defined by $M(R_{\rm DS})=M_{\rm DS}$, with $M_{\rm DS}$ the total mass of the DS. We also define the compactness $C=M_{\rm DS}/R_{\rm DS}$, which is a relevant quantity for the GW signal, as will be discussed in section~\ref{sec:GW_signals}. Finally, after determining the pressure distribution with the TOV equation, we determine the density distribution using the EoS. 

\begin{figure}[t!]
	\centering
	\includegraphics[width=.48\textwidth]{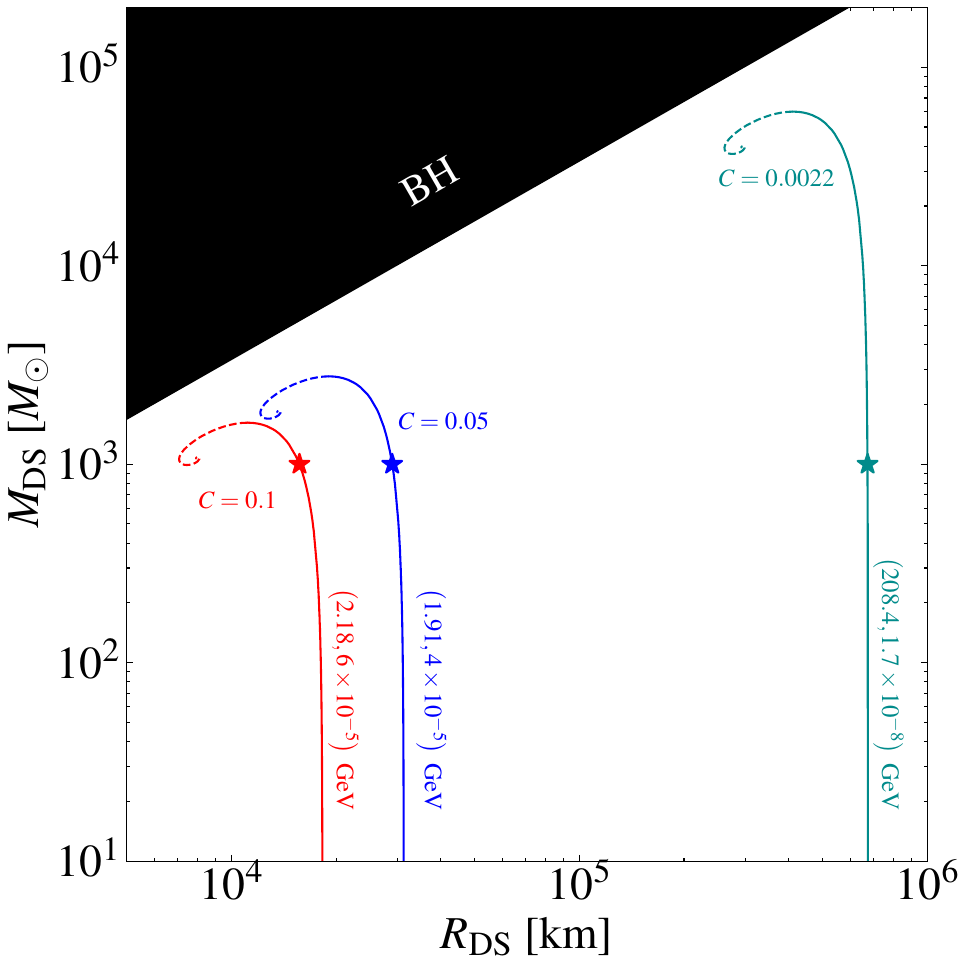}
	\includegraphics[width=.48\textwidth]{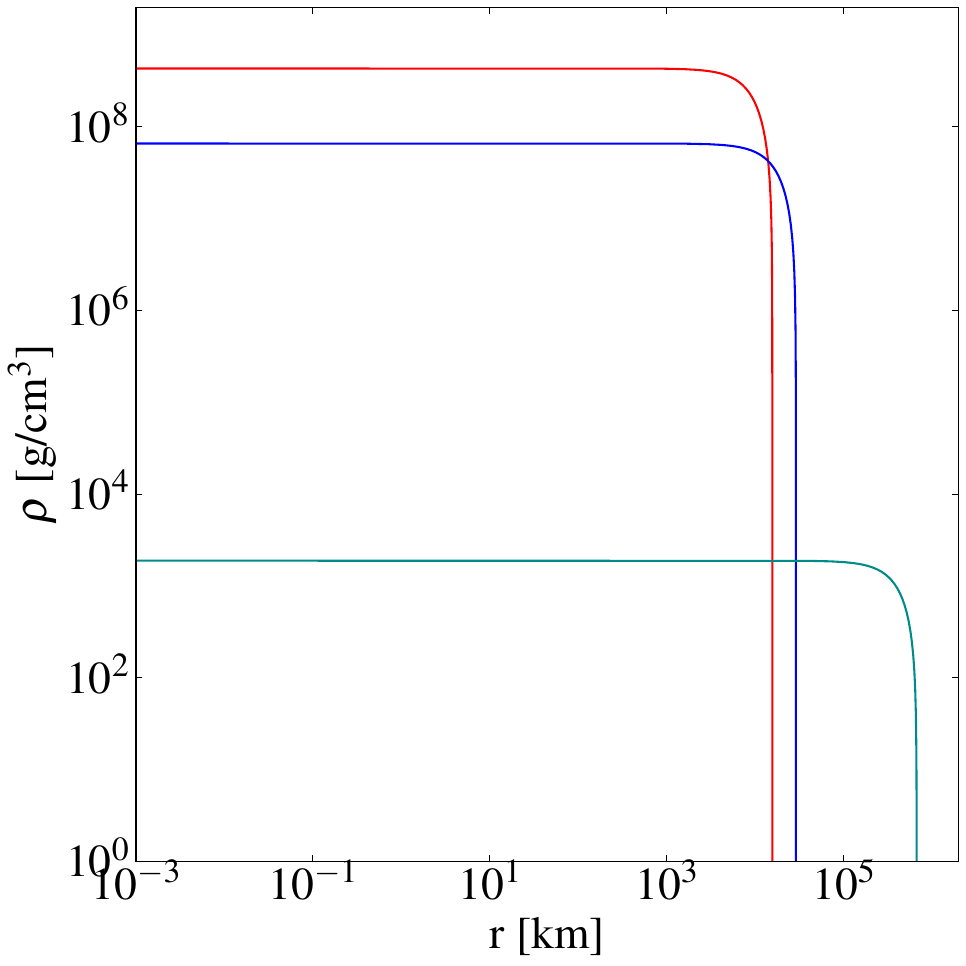}
	\caption{\textit{Left panel:} Mass-radius relation for a DS with EoS~\eqref{eq:EOS} for $\alpha_D=0.1$ and three choices of dark electron and dark photon mass. The star-shaped points correspond to configurations of DS mass of $1000M_\odot$, but with different compactness. The black-colored area shows the unavailable region of compactness $C\geq0.5$, corresponding to BHs. \textit{Right panel:} Mass density distribution as a function of the radius of the DS. The density distributions correspond to the benchmark points (colored stars) indicated in the left panel.}
	\label{fig:mass_vs_r}
\end{figure}

In the left panel of Fig.~\ref{fig:mass_vs_r}, we show the mass-radius relation of asymmetric DSs for the benchmark parameters $m_{e_D}=2.18~\rm GeV$ and $m_{\gamma_D}=60~\rm keV$ (red), $m_{e_D}=1.91~\rm GeV$ and $m_{\gamma_D}=40~\rm keV$ (blue) and $m_{e_D}=208.4~\rm GeV$, and $m_{\gamma_D}=17~\rm eV$ (cyan), with a strength of the self-interaction in all three cases $\alpha_D=0.1$. Each of the mass-radius curves has a maximum value for the asymmetric DS mass, in analogy to the ``Chandrasekhar mass" of white dwarfs, and only the branch of the curve with increasing radius from the maximum leads (solid curves) to stable configurations, while the opposite branch leads to unstable solutions (dashed curves). The black area denotes the region where the radius becomes smaller than the Schwarzschild radius and therefore we have BH formation. In the right panel we show the density distribution inside the DS for the three benchmark points indicated in the left panel.

\begin{figure}[t!]
	\centering
	\includegraphics[width=.58\textwidth]{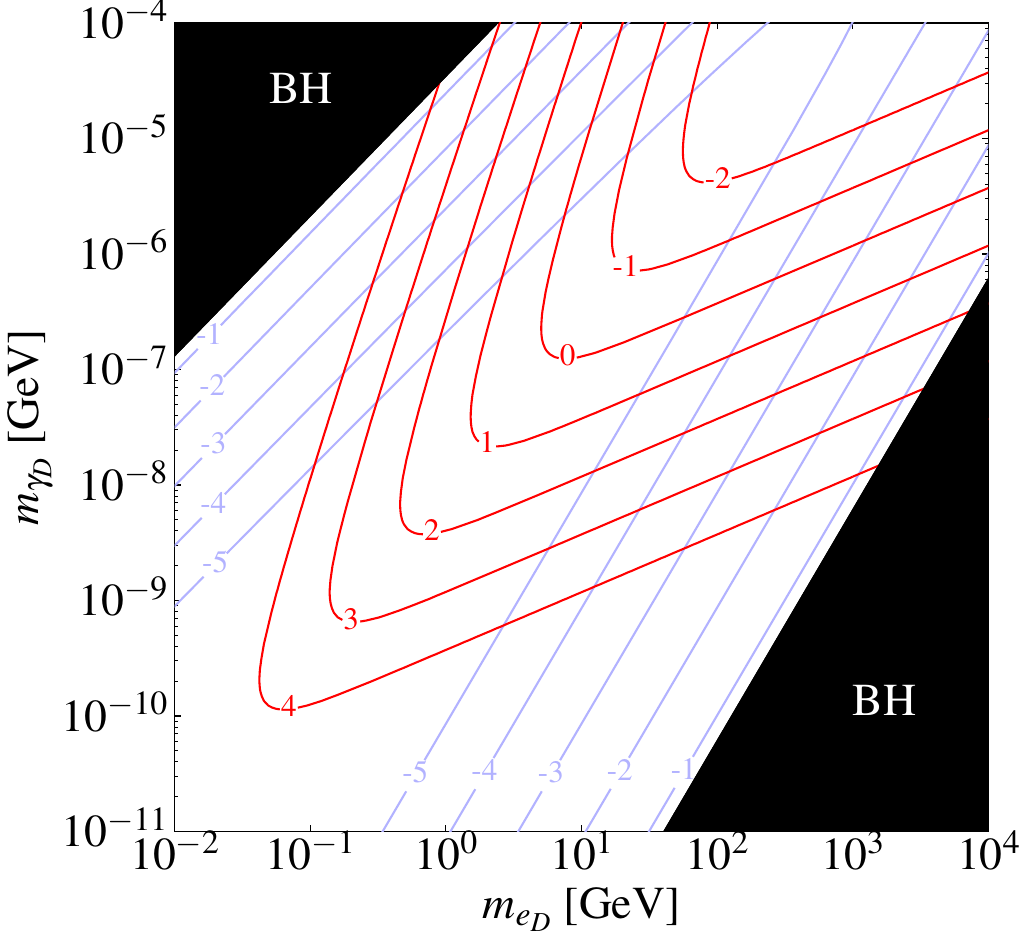}
	\caption{Parameter space in dark photon vs dark electron mass for fixed $\alpha_D=0.1$. The red contours correspond to constant DS mass (in units of $\log_{10}(M_{\rm DS}/M_\odot)$), while the light blue ones show the compactness in units of $\log_{10}(C)$. The black-shaded regions are the space where the solutions satisfy $C\geq0.5$, and correspond to BHs.}
	\label{fig:MassCompactnessParameterSpace}
\end{figure}

From the left panel of Fig.~\ref{fig:mass_vs_r}, we see that for a fixed set of microphysical parameters, there exists a family of solutions of DSs. They correspond to different choices of the core density (or pressure). On the other hand, we note that phase-space trajectories of the dark electron clump shown in Fig.~\ref{fig:Formation_phase_space} predict a final fragment with a unique mass, determined entirely by the microphysics and not the particular halo mass. By combining both results, namely the mass-radius relation and the unique mass of the final fragment from formation, we are able to pick the particular solution of DS and obtain both its radius and compactness. Hence, for a given choice of $\{m_{e_D},m_{\gamma_D},\alpha_D\}$ we are able to determine the unique DS mass, radius and compactness. In Fig.~\ref{fig:MassCompactnessParameterSpace}, we present the parameter space in dark electron vs dark photon mass for fixed $\alpha_D=0.1$. The red contours correspond to the DS mass in units of $\log_{10}(M_{\rm DS}/M_\odot)$ and the light blue contours show the compactness in units of $\log_{10}(C)$. The black-shaded regions show unphysical solution with $C>0.5$, which are impossible to obtain with the mass-radius relation. 

We note that, while it is possible to obtain DSs with the same mass from different sets of parameters, most solutions will have different radii. Nevertheless, we can see from  Fig.~\ref{fig:MassCompactnessParameterSpace} that we are able to derive the same DS solution (same mass and radius) for two sets of parameters. For example, for a DS with a mass of $10^4M_\odot$ and compactness of $C=10^{-5}$, we have two sets: $\{m_{e_D}\sim0.1~\mathrm{GeV},m_{\gamma_D}\sim10~\mathrm{eV}\}$ and $\{m_{e_D}\sim4~\mathrm{GeV},m_{\gamma_D}\sim0.5~\mathrm{eV}\}$. While both choices will lead to the same structure, the expected cosmological DS mass density will be different, as the predicted minimal and maximal halo masses (see Eqs.~\eqref{eq:M_halo_min_def} and~\eqref{eq:max_halo_mass}) will change with different dark electron and dark photon mass choices.

\subsection{Merger cross section}\label{sec:merger_cross_section}
When two compact objects interact through gravitational focusing, their trajectories may lead either to a direct collision or to a bound merger driven by the loss of orbital energy through GWs and tidal dissipation. The rate at which such mergers occur depends on the corresponding cross section, which we compute below.

The total energy loss during a close encounter can be written as the sum of the gravitational-wave (GW) and tidal contributions (tide),
\begin{equation}
    \Delta E_{\rm loss} = \Delta E_{\rm GW} + \Delta E_{\rm tide}.
\end{equation}
For a parabolic encounter, the GW energy loss is given by~\cite{1977ApJ...216..610T,Mouri:2002mc}
\begin{equation}
    \Delta E_{\rm GW} = \frac{85\pi G^{\frac{7}{2}}}{12\sqrt{2}}
    \frac{M_1^2 M_2^2(M_1+M_2)^{1/2}}{R_p^{7/2}},
\end{equation}
where $M_1$ and $M_2$ are the stellar masses and $R_p$ is the pericenter distance of the encounter.

The tidal energy deposited in the stellar oscillation modes during the passage is~\cite{1977ApJ...213..183P,Hoang:2020gsi}
\begin{equation}\label{eq:DeltaE_sum}
    \Delta E_{\rm tide} = 
    \frac{GM_{2}^{2}}{R_1}\sum_{\ell=2}^{3}
    \left(\frac{R_1}{R_p}\right)^{2\ell+2}T_{\ell,1}(\eta_1)
    +\frac{GM_{1}^{2}}{R_2}\sum_{\ell=2}^{3}
    \left(\frac{R_2}{R_p}\right)^{2\ell+2}T_{\ell,2}(\eta_2),
\end{equation}
where $R_1$ and $R_2$ are the stellar radii and $\eta_i=\sqrt{M_i/(M_i+M_{j\neq i})}(R_p/R_i)^{3/2}$ is a dimensionless parameter related to the duration of a periastrion passage relative to the hydrodynamical timescale of star $i$. The function $T_{\ell,i}$ is the energy-contribution fraction of star $i$, where the index $\ell$ indicates the harmonic mode ($\ell=2$ corresponds to quadrupole and $\ell=3$ to octupole). The $\ell=2,3$ terms give a $99\%$ contribution to the dissipated energy, hence we consider no higher order harmonic modes. Details on the calculation of these functions are given in appendix~\ref{app:tidal_loss}.

The pericenter distance $R_p$ corresponding to a marginally bound system is determined by equating the total energy loss to the initial relative kinetic energy,
\begin{equation}
    \Delta E_{\rm loss} = \frac{1}{2}\mu v_{\rm rel}^2,
\end{equation}
where $\mu=M_1M_2/(M_1+M_2)$ is the reduced mass and $v_{\rm rel}$ is the initial relative velocity at infinity. Solving this relation yields the critical $R_p$ for which a capture (or merger) occurs. The corresponding gravitational focusing cross section for merger is then
\begin{equation}\label{eq:cross_section}
    \sigma_m = \frac{2\pi G(M_1+M_2)R_p}{v_{\rm rel}^2},
\end{equation}
where we recover the well-known formula for the two-body BH scattering cross-section for
GW emission by setting $\Delta E_{\rm tide}=0$
\begin{equation}\label{eq:sigma_GW}
    \sigma_{\rm GW}=2\pi\left(\frac{85\pi}{6\sqrt{2}}\right)^{\frac27}G^2\frac{(M_1+M_2)^{\frac{10}{7}}M_1^{\frac27}M_2^{\frac27}}{v_{\rm rel}^{\frac{18}{7}}}.
\end{equation}
On the other hand, the geometrical (collisional) cross section is
\begin{equation}\label{eq:collisional}
    \sigma_{\rm coll} = \pi (R_1 + R_2)^2
    \left[1 + \frac{2G(M_1 + M_2)}{(R_1 + R_2)v_{\rm rel}^2}\right].
\end{equation}
The effective cross section governing the interaction rate is taken as the maximum of these two values,
\begin{equation}\label{eq:effective_cross_section}
    \sigma_{\rm merge} = \max(0,\sigma_m- \sigma_{\rm coll}).
\end{equation}

\begin{figure}[t!]%
\centering
\includegraphics[width=.58\textwidth]{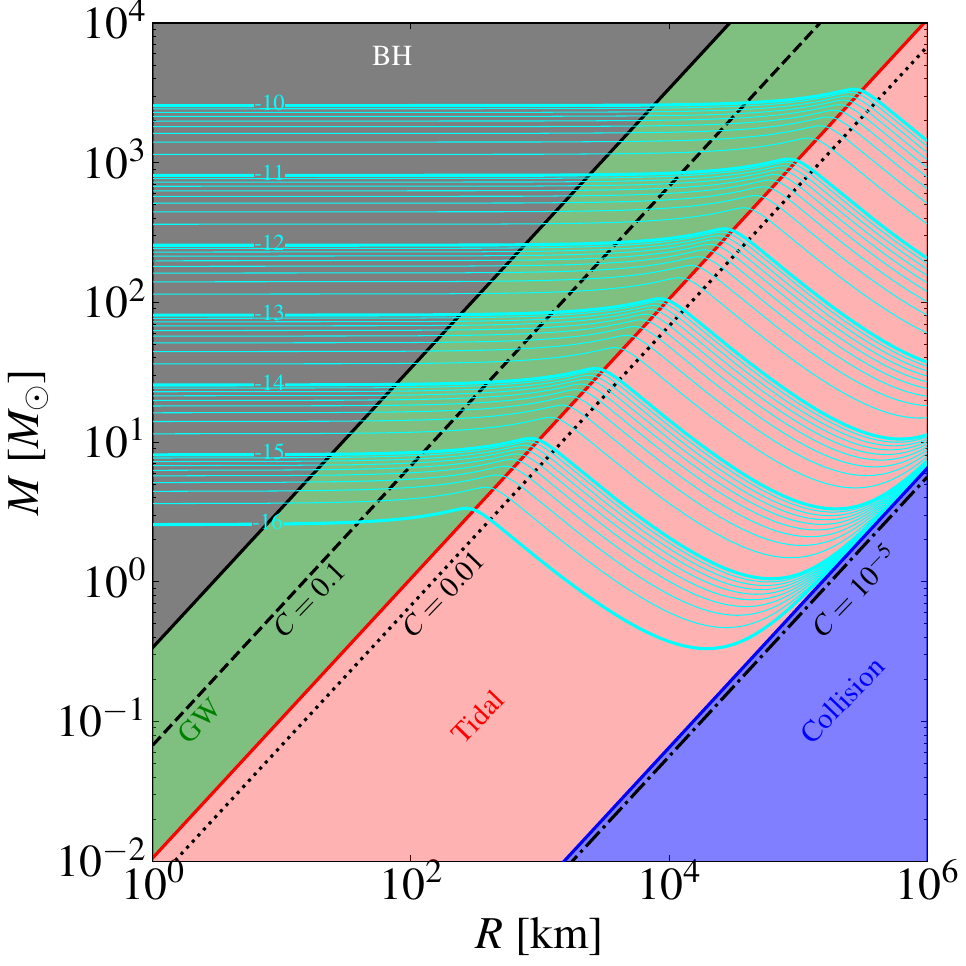}
\caption{Merger cross section contours (cyan, in units of $\log_{10}(\sigma_{\rm merge}/{\rm pc}^2)$) in the DS mass vs radius plane for $v_{\rm rel}=200~\mathrm{km}~\mathrm{s}^{-1}$. The black lines correspond to contours of constant compactness and the black region correspond to BH. The green area shows the region where tidal capture and collision are subdominant and the cross section is given by~\eqref{eq:sigma_GW}, the red area is where tidal effects dominate and the blue area where the geometrical cross section is most important. }
\label{fig:CrossSection}
\end{figure}

In Fig.~\ref{fig:CrossSection}, we depict the processes that dominate the merger cross section in the DS mass-radius parameter space using different colors. The black curves represent constant compactness, while the black region corresponds to BH. The green region, labeled as ``GW'', shows the space where tidal deposition and collision are subdominant and the merger cross-section is dominated by the standard two-body BH scattering cross-section given in Eq.~\eqref{eq:sigma_GW}. Instead, in the red region  tidal effects constitute the dominant energy loss mechanism for the binary. Finally the blue region is where the collisional cross-section is greater than the merger cross-section. The cyan curves are contours of constant merger cross-section in units of $\log_{10}(\sigma_{\rm merge}/{\rm pc}^2)$. We note that, for large compactness ($C\gtrsim0.01$), DS collision is similar to the BH scenario and there is no enhancement due to their deformability. On the other hand, for moderate compactness $10^{-5}<C<10^{-2}$, tidal energy loss is dominant and leads to an increase in the merger cross-section. Finally, for $C<10^{-5}$, the size of the DSs is so large that their direct collision is more likely than the formation of a bound system. Such events, while likely to emit large-amplitude GWs, are probably outside the sensitivity of detectors and are not well studied as typical mergers. Therefore in order to be on the conservative side, we will not consider  their GW contribution further.

\section{Gravitational wave signals from dark star mergers}\label{sec:GW_signals}

%\subsection{Roche limit}
%The calculation of the total merger rate must account for regions where DSs cannot survive. Within the tidal radius of the central massive black hole, DSs are disrupted and cannot merge. This tidal radius is given by:

%\begin{equation}
%    r_{\rm t}=2^{\frac13}R_{\rm DS}\left(\frac{M_{\rm BH}}{M_{\rm DS}}\right)^{\frac13}\approx1260~\mathrm{km}\left(\frac{R_{\rm DS}}{10~\rm km}\right)\left(\frac{M_{\rm BH}}{10^6M_\odot}\right)^{\frac13}\left(\frac{M_{\rm DS}}{M_\odot}\right)^{-\frac13}.
%\end{equation}
%Therefore, the lower limit for the spatial integral is set by the more restrictive of these two conditions:
%\begin{equation}
%    N_{\rm halo}=\int\displaylimits_{\max\{r_{\rm t},4r_{\rm sch}\}}^{r_{\rm vir}}\frac{1}{2}\left(\frac{f_{\rm DS}\rho_{\rm DM}(r)}{M_{\rm DS}}\right)^2 \sigma_{\rm merge}(r) v_{\rm rel}(r) d^3r.
%\end{equation}

\subsection{Dark star merger rate}

The formation of DS binaries in DM overdense regions can be achieved via gravitational capture like scattering and gravitational bremsstrahlung emission \cite{Bird:2016dcv}. Under certain conditions, when two DSs encounter each other, rapid energy loss via emission of GWs leads to formation of a gravitational bound DS binary. Once it forms, GW emission from this BDS would drive its merger. The capture timescale is much longer than the merger timescale of BDSs. Hence, we can use the capture rate of DSs as an estimate of the merger rate. In the following discussion, we focus on the DS merger rate contribution from two-body capture rather than  other mechanisms like a three-body interaction, since the two-body capture produces more BDSs than the others when the energy density fraction of DS in DM is much smaller than one \cite{Franciolini:2022ewd}.

The DS merger rate depends on the DS number density, the DS merger cross section, and the relative velocity of the two DSs. The DS number density can be estimated by the energy density and mass of DSs as $n_{\rm DS} = \rho_{\rm DS}/M_{\rm DS} = f_{\rm DS}\rho_{\rm DM}/M_{\rm DS}$, where $f_{\rm DS}$ is the energy density fraction of DS in DM. Hence, a larger DM density would contribute significantly to the DS merger rate. The DM density in a DM halo is several orders of magnitude larger than the average DM density in the Universe, following potentially an NFW profile as in Eq.~\eqref{eq:NFW_profile}. The presence of an SMBH in the halo triggers the formation of a DM spike around it, making the DM density profile even steeper as seen in Eq.~\eqref{eq:spike_profile}. Since the  density in the spike is much larger than the inner NFW density, the DS merger rate in the spike would dominate the total DS merger rate in the observations \cite{Nishikawa:2017chy,Ding:2024mro}. The merger rate of DSs in the DM spike can be estimated as follows
\begin{equation}\label{eq:merger_rate_spike}
    N_{\rm sp}=\int\displaylimits_{\max\{r_{\rm t},4R_{s}\}}^{r_{\rm sp}}\frac{1}{2}\left(\frac{f_{\rm DS}\rho_{\rm sp}(r)}{M_{\rm DS}}\right)^2 \sigma_{\rm merge}(r) v_{\rm rel}(r) d^3r~.
\end{equation}
Here, the energy density fraction of DSs in DM $f_{\rm DS}$  depends on two factors: the fraction of DM composed of dark electrons $f_{e_D}$, and the fraction of those dark electrons that have condensed into compact DSs $f_{e_D-\rm DS}$ (fixed here to $0.5$) and therefore $f_{\rm DS}=f_{e_D}f_{e_D-\rm DS}$.\footnote{Note that $f_{\rm DS}$ is linearly proportional to the product $f_{e_D}f_{e_D-\rm DS}$. Consequently, varying $f_{e_D-\rm DS}$ is degenerate with varying $f_{e_D}$ with respect to the final value of $f_{\rm DS}$.} Adopting \textit{e.g.} $f_{e_D} = 10 \%$ yields $f_{\rm DS} \sim 5\%$ in the present Universe. The merger cross section has been  discussed in Sec.~\ref{sec:merger_cross_section}. The relative velocity can be estimated as a circular velocity around the SMBH at a given radius $r$ as
\begin{equation}
    v_{\rm vel} = \sqrt{\frac{G M_{\rm BH}}{r}}~.
\end{equation}
The upper bound of  integration is the radius of the spike that follows from Eq.~\eqref{eq:rsp}, and the lower bound of integration is the maximum of the inner DM spike radius $4 R_{s}$ ($R_s$ being the SMBH Schwarzschild radius) and the Roche limit radius $r_{\rm t}$. The  Roche limit $r_{\rm t}$ characterizes the radius from the SMBH below which DSs are disrupted by tidal effects and do not survive and it is given by~\cite{1982S&T....64Q.152S}
\begin{equation}
    r_{\rm t}=2^{\frac13}R_{\rm DS}\left(\frac{M_{\rm BH}}{M_{\rm DS}}\right)^{\frac13}\approx1260~\mathrm{km}\left(\frac{R_{\rm DS}}{10~\rm km}\right)\left(\frac{M_{\rm BH}}{10^6M_\odot}\right)^{\frac13}\left(\frac{M_{\rm DS}}{M_\odot}\right)^{-\frac13}.
\end{equation}
For a DS with $M_{\rm DS}= \, M_\odot$  and $R_{\rm DS}=10 \, {\rm km}$  rotating around an SMBH with a mass of $10^{6}\, M_\odot$, 
$r_{\rm t}\simeq 1260\, {\rm km}$. 

\begin{figure}[t!]
	\centering
	\includegraphics[width=.48\textwidth]{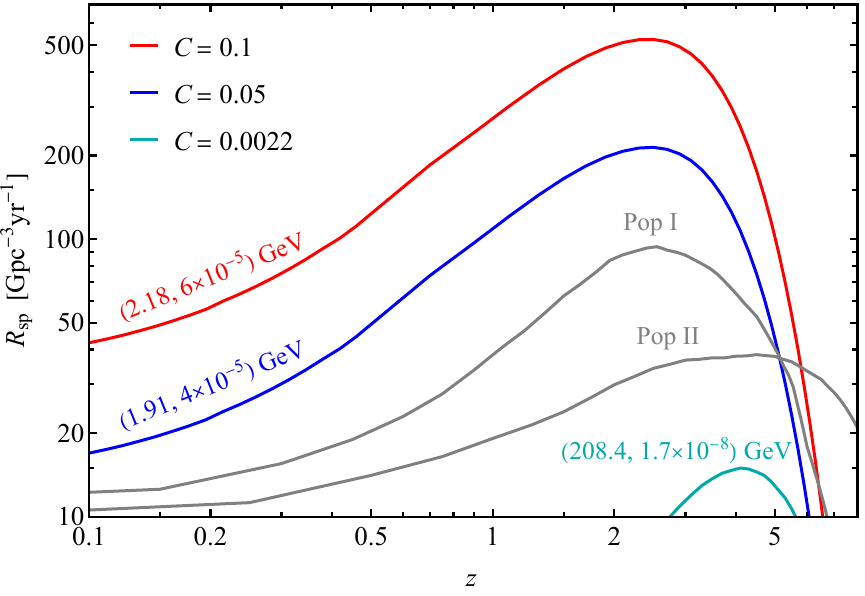}
    \includegraphics[width=.48\textwidth]{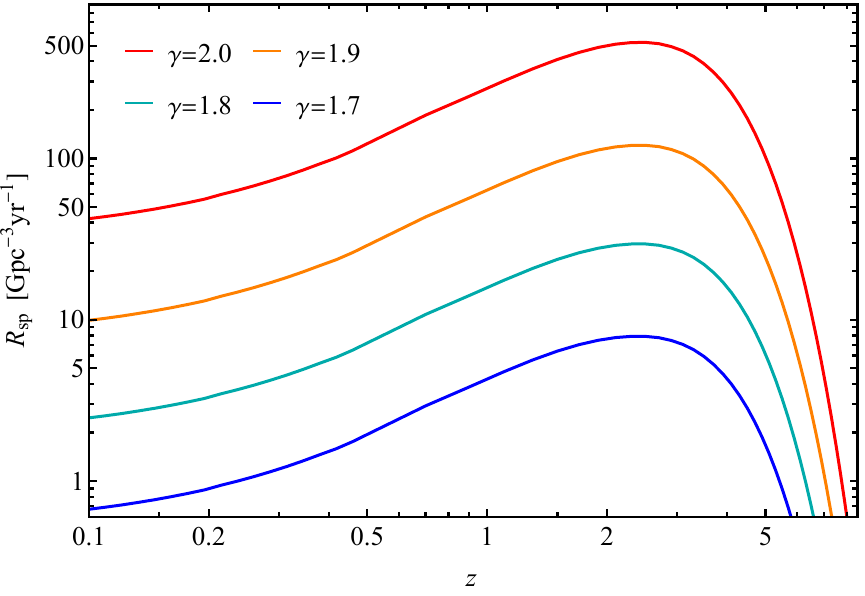}
	\caption{The redshift dependence of the DS merger rate in the DM spike. \textit{Left panel}: The  merger rate for three selected benchmark sets of parameters $(m_{e_D}, m_{\gamma_D}) = (2.18, 6\times10^{-5}) \, \text{GeV}, \,(1.91, 4\times10^{-5}) \, \text{GeV}, \, (208.4, 1.7\times10^{-8}) \, \text{GeV}$. These three benchmark parameters produce the same DS mass of $1000\,M_\odot$ and different compactness of $C=0.1,\,0.05,\,0.0022$, respectively. The DS merger rate is estimated with a power index $\gamma = 2$. The gray curves are the merger rates of Pop I and II BHs obtained from Ref.~\cite{Tanikawa:2021qqi}. \textit{Right panel}: The DS merger rate with different power index $\gamma = 2.0,\,1.9,\,1.8,\,1.7$. The DS is produced with parameters of $(m_{e_D}, m_{\gamma_D}) = (2.18, 6\times10^{-5}) \, \text{GeV}$, which determines a $1000\, M_\odot$ DS mass and $0.1$ compactness.}
	\label{fig:mergerrate_redshift}
\end{figure}

Eq.~\eqref{eq:merger_rate_spike} gives the DS merger rate in one DM spike $N_{\rm sp}$, and the most important factor in $N_{\rm sp}$ is the density profile of the DM spike, which varies for different SMBH mass $M_{\rm BH}$ and inner DM halo power index $\gamma$. For a fixed value of $\gamma$, $N_{\rm sp}$ is a function of the SMBH mass.
To obtain the total DS merger rate, we should add the contribution of all spikes by taking the convolution of Eq.~\eqref{eq:merger_rate_spike} with the SMBH mass function 
\begin{equation}\label{eq:merger_rate}
    R_{\rm sp} =\int_{M_{\rm BH,min}}^{M_{\rm BH,max}} N_{\rm sp}(M_{\rm BH}) \frac{d n}{d M_{\rm BH}} dM_{\rm BH}~,
\end{equation}
where $M_{\rm BH, min}$ and $M_{\rm BH,max}$ are respectively the minimum and  maximum SMBH masses. In general, to form a DM spike around an SMBH, a minimum SMBH mass $M_{\rm BH, min} = 10^5 - 10^6\,M_\odot$ is needed. Furthermore the maximum SMBH mass is $M_{\rm BH, max} = 10^9-10^{10}\,M_\odot$. However we must ensure that DSs are able to form in the host DM halo of this SMBH mass range. Therefore, we set $M_{\rm BH,min}$ to be  the maximum  of $10^5\, M_\odot$ and the minimal SMBH mass whose corresponding DM halo can produce DSs, and we set $M_{\rm BH, max}$ to be the minimum of $10^{10} \,M_\odot$ and the maximum SMBH mass whose corresponding DM halo can produce DSs.
Then we can use Eq.~\eqref{eq:merger_rate} to calculate the redshift evolution of DS merger rate as shown in Fig.~\ref{fig:mergerrate_redshift}.

Figure~\ref{fig:mergerrate_redshift} (left panel) shows the redshift dependence of the DS merger rate computed
for three benchmark microphysical parameter choices that produce the same DS mass $M_{\rm DS}=1000\,M_\odot$ but with different compactness. The most compact configuration ($C=0.1$) yields the largest merger rate, exceeding the Pop I/II BH rate by up to a factor $\sim 5$, the intermediate compactness ($C=0.05$) produces an enhancement by a factor of~$\sim 2$, while the least compact DS ($C\sim 0.002$) results in a merger rate below the Pop I/II BH contribution. This may seem contradictory. As we have argued in Sec.~\ref{sec:merger_cross_section}, for smaller compactness, the merger cross section increases, which would naively mean  a larger merger rate according to Eq.~\eqref{eq:merger_rate_spike}. However, we point out that the merger rate does not depend solely on the cross section but on other factors too.

Firstly, Eq.~\eqref{eq:merger_rate_spike} depends explicitly on $r_{\min}=\max\{r_t,4R_{s}\}$. Because the spike density $\rho_{\rm sp}(r)$ rises sharply toward the SMBH, the integrand is dominated by the innermost radii. Therefore, any mechanism that effectively increases $r_{\min}$ (and removes the inner spike) strongly suppresses the total $N_{\rm sp}$. For a fixed $M_{\rm DS}$ decreasing compactness corresponds to an increased
stellar radius $R_{\rm DS}$ and therefore to a larger tidal radius, moving $r_{\min}$ to larger values and removing DSs from the highest-density region of the spike. Hence, small-compactness DSs lead to strong suppression of the merger rate despite larger capture cross sections.

Secondly, although smaller compactness increase the capture cross section, too small compactness makes the cross section to be dominated by direct collisions
as shown in Fig.~\ref{fig:CrossSection} which potentially produce GW signals outside the detector range.

Thirdly, for  moderate/low compactness ($10^{-5}\lesssim C\lesssim 10^{-2}$)  where the cross section is dominated by tidal energy losses and indeed the capture cross section is larger than that of compact BDS, the merger rate of compact BDS can still be larger than that of lower compactness BDS because the DS density in the former can be larger than that of the latter. Note that Fig.~\ref{fig:DSMD} reveals two facts. The first is that DS densities increase as we lower the redshift. The second is that for low redshift, higher compactness DSs have higher densities compared to DSs with similar mass and lower compactness. This is related to the fact that different DM parameters lead to different DS formation conditions in halos. Because $N_{\rm sp}\propto (n_{\rm DS})^2\propto (\rho_{\rm DS}/M_{\rm DS})^2$, smaller cosmological abundances at small redshift strongly reduce the merger rate. 

Fourthly, apart from the DS density that is affected by redshift, another source of dependence on redshift for the merger rate is the SMBH mass function in Eq.~\eqref{eq:merger_rate}, which counteracts the behavior of the DS density. Rapid SMBH formation at high redshift contributes a large value of $dn/dM_{\rm BH}$ in Eq.~\eqref{eq:merger_rate} and it is the interplay of the two factors that produce a peak of DS merger rate at redshift around $z \sim 2-4$. 

These five effects (merger cross section, direct collisions, tidal radius, cosmological abundance, and the SMBH mass function)  explain  the  observed pattern in the left panel of Fig.~\ref{fig:mergerrate_redshift}. The potential statistical study of different BDS mergers at different redshifts might offer a unique perspective to unveil DS properties.

The right panel of Figure~\ref{fig:mergerrate_redshift} depicts the redshift dependence of the DS merger rate for $\gamma = 1.7,\,1.8,\,1.9,\,2.0$. In general, a larger $\gamma$  enhances the DM density in the spike and decreases the spike radius. Since the inner part of the spike gives the dominant contribution to the  DS merger rate, a smaller  spike radius doesn't change the merger rate too much. Hence, a larger $\gamma$ results in a larger  merger rate due to a larger DM density in the spike, as illustrated in the right panel of Fig.~\ref{fig:mergerrate_redshift}. 
%Meanwhile the power index $\gamma$ describes the inner density profile of DM halo, which can help unveil the distribution of DM in the galactic center.

\subsection{Detectability}

\begin{figure}[t!]
	\centering
	\includegraphics[width=.48\textwidth]{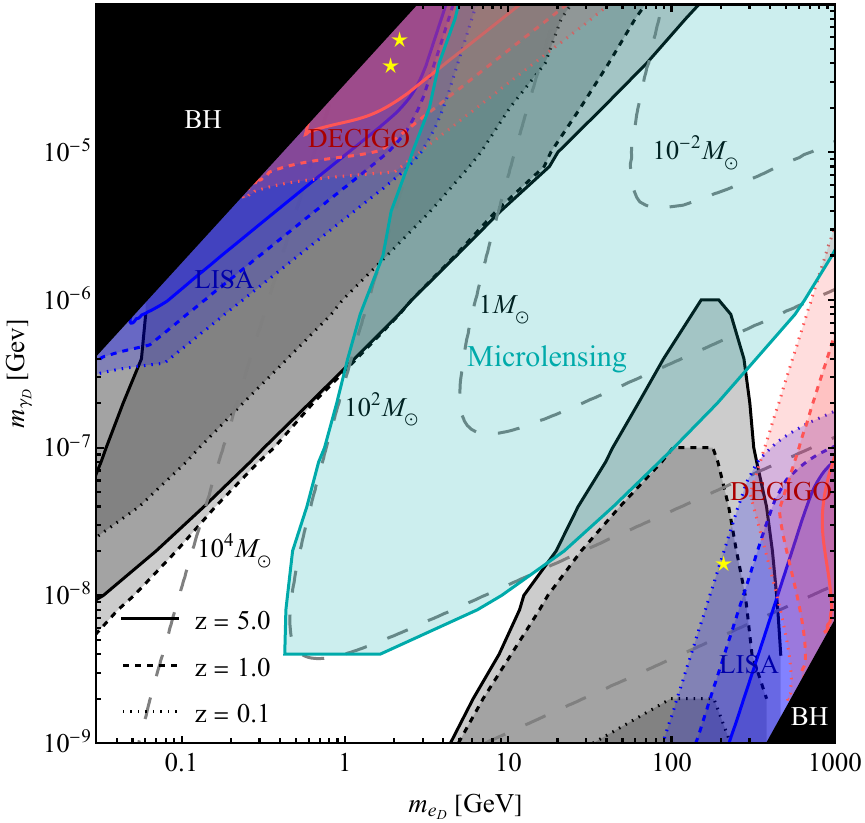}
        \includegraphics[width=.48\textwidth]{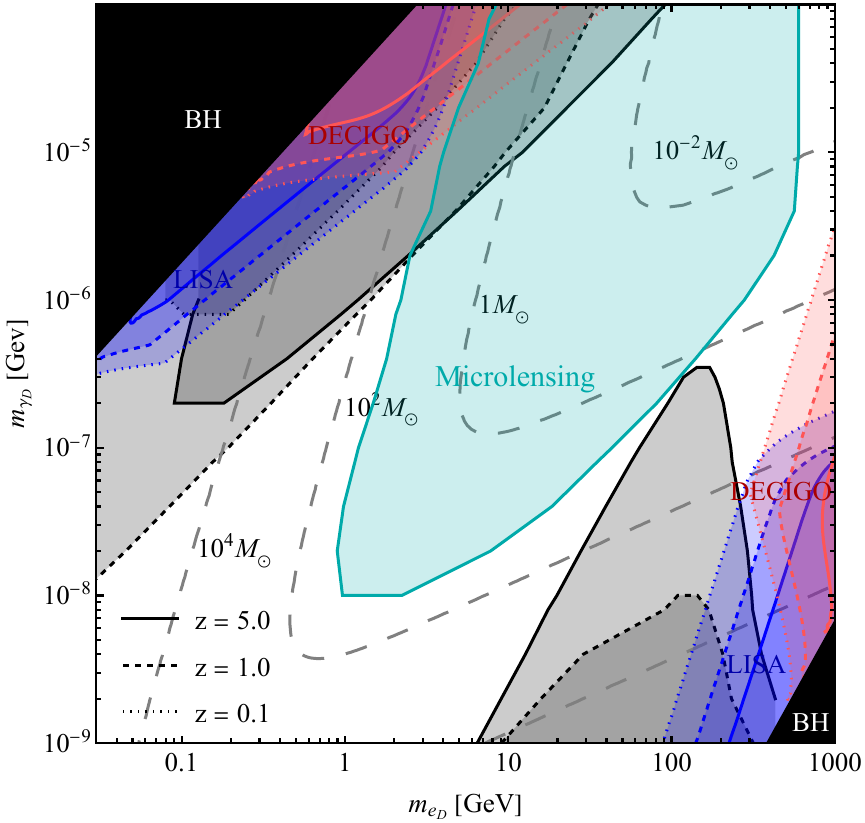}
        \includegraphics[width=.48\textwidth]{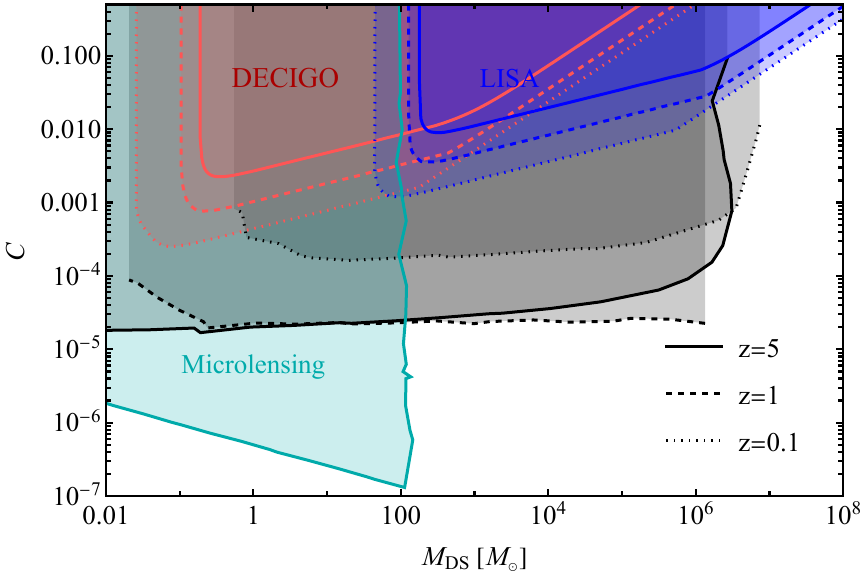}
        \includegraphics[width=.48\textwidth]{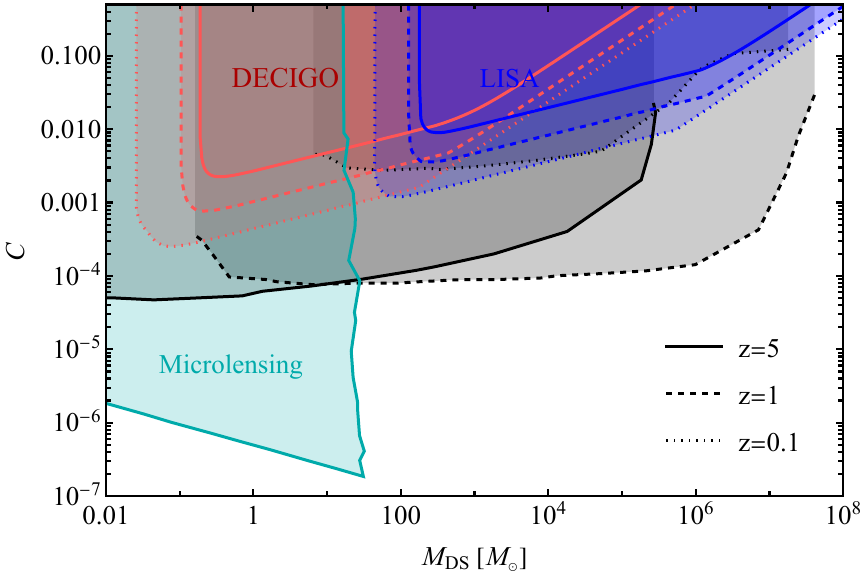}
	\caption{The detectable parameter regions of DS mergers. The gray  regions depict DS merger rate $R>1\,{\rm Gpc^{-1}\,yr^{-1}}$. The red and blue  regions represent the detectable parameter regions by the GW detectors DECIGO and LISA, respectively. Enclosed by the solid, dashed and dotted curves denote regions where the aforementioned merger rate and detectability of DSs are achieved within  redshift $z=5.0,\,1.0,\,0.1$, respectively.  The cyan  region is the parameter region ruled out  by  microlensing observations. \textit{Upper left} and \textit{upper right} panels show the detectable parameter regions of $m_{e_D}$ and $m_{\gamma_D}$ with  $f_{e_D} = 10\%$ (left) and $f_{e_D}=5\%$ (right). The gray dashed curves correspond to the DM mass contour of $M_{\rm DS} = 10^{-2},\,1,\,10^2,\,10^4\, M_\odot$. Three yellow stars label the benchmark points with parameters of $(m_{e_D}, m_{\gamma_D}) = (2.18, 6\times10^{-5}) \, \text{GeV}, \,(1.91, 4\times10^{-5}) \, \text{GeV}, \, (208.4, 1.7\times10^{-8}) \, \text{GeV}$. The black regions correspond to BH formation since compactness reaches $C = 0.5$. \textit{Lower left} and \textit{lower right} panels show the detectable parameter regions of DS mass $M_{\rm DS}$ and compactness $C$ again for the two distinct values of $f_{e_D}$. }
	\label{fig:detectability}
\end{figure}

In order to observe the GW signals from DS mergers, one important requirement is that their GW should be detectable during the merger process. Depending on the DM parameters $m_{\gamma_D}$, $m_{e_D}$ and their coupling constant $\alpha_D$, various parameters produce distinct DS mass and compactness. Among the whole  parameter space, it is only part of it that can produce detectable DS mergers. The detection of GWs from  BDSs requires a signal-to-noise ratio SNR larger than a conservative threshold value. The SNR is defined as \cite{Rosado:2015voo, Ding:2020ykt}
\begin{equation}
    \textrm{SNR} = \sqrt{4\int_{f_{\rm min}}^{f
    _{\rm max}}\frac{|\tilde{h}(f)|^2}{S_n(f)} df}~,
\end{equation}
where $S_n (f)$ is the noise strain of the GW detector \cite{Moore:2014lga}  (we consider LISA and DECIGO in our calculation), and $h(f)$ is the Fourier transform of the GW waveform $h(t)$ \cite{Droz:1999qx}, which can be expressed as
\begin{equation}\label{eq:h_f}
    \tilde{h}(f) \simeq \sqrt{\frac{5}{24}} \frac{(G \mathcal{M}_c (1+z))^{5/6}}{\pi^{2/3} d_L(z)} f^{-7/6}~,
\end{equation}
where $\mathcal{M}_c \equiv M_{\rm DS}/2^{1/5}$ is the chirp mass of equal mass BDSs, and $d_L(z)$ is the luminosity distance between the observers and the BDSs.  This formula describes the GW waveform from BBHs, which can approximate the GW waveform of DSs for frequency smaller than $f_{\rm ISCO} = C^{3/2}/(6\sqrt{3} \pi GM_{\rm DS})$ \cite{Giudice:2016zpa}.
$f_{\rm max}$ is the maximum frequency of BDSs in the detector frame, where we use $ f_{\rm max} = f_{\rm ISCO}/(1+z)$ in Eq.~\eqref{eq:h_f} as an approximation of the GW waveform of the BDSs. $f_{\rm min}$ is the minimum frequency of BDSs, and it is set by the maximum of the detector's lower frequency bound in LISA and DECIGO and the GW frequency of BDSs evolved backwards in time for the duration of an observation time from the maximal frequency $f_{\rm max}$. In what follows we set this observation time to one year. This backward evolving frequency in the detector frame is given by
\begin{equation}
    \frac{df}{dt} =\frac{96}{5} [G \mathcal{M}_c (1+z)]^{5/3} \pi^{8/3} f^{11/3}~.
\end{equation}
To ensure a detection probability greater than $95\%$, corresponding to a false-alarm probability below $0.1\% $, the SNR must exceed a conservative threshold value that can be taken to be \cite{LIGOScientific:2016vbw}
\begin{align}
    {\rm SNR} > 8~.
\label{SNR8}
\end{align}
Then we can obtain the detectable parameter regions for DS mergers in the two GW detectors LISA and DECIGO as shown in Fig.~\ref{fig:detectability}.

To detect the merger of a DS binary, we not only require that its GW signal can be probed by GW detectors, but also that its merger rate in the DM spike should at least reach a conservative value at various redshifts to make sure the redshift evolution of BDS mergers, as illustrated in Fig.~\ref{fig:mergerrate_redshift}, can be detected within the duty time of GW detectors. Hence we require
\begin{equation}
    R_{\rm sp} > 1\, {\rm Gpc^{-3} yr^{-1}}~.
\end{equation}
This detection requirement along with the LISA and DECIGO detection criteria would constrain the detectable DS parameter region in Fig.~\ref{fig:detectability}.

In addition, astrophysical observations impose constraints on the parameters of DSs. In particular, microlensing observations provide an upper bound on the mass density fraction $f_{\rm MACHO}(M)$ of massive compact halo objects (MACHOs) in the DM abundance. Recent limits have been summarized and provided in Fig.~3.3 of Ref.~\cite{Cirelli:2024ssz}. Hence, the allowed energy density fraction of DSs in DM should satisfy
\begin{equation}\label{eq:microlensing}
    f_{\rm DS}(\Theta) < f_{\rm MACHO}(M_{\rm DS}),
\end{equation}
where $\Theta$ is the DS parameter set, such as $m_{e_D}$, $m_{\gamma_D}$, $\alpha_D$, etc. A parameter set  determines the DS mass $M_{\rm DS}$ and the energy density fraction $f_{\rm DS}$. In order to avoid the microlensing constraints, this fraction should be smaller than the upper bound of a MACHO with mass $M_{\rm DS}$ as described in Eq.~\eqref{eq:microlensing}.

It is also worth noting that recent results from pulsar timing arrays (PTAs), such as NANOGrav~\cite{NANOGrav:2023gor}, have been interpreted as providing constraints on populations of compact objects in a similar mass range (mainly $M\lesssim M_\odot$). However, these constraints do not apply to the DS scenario considered here. PTA constraints on MACHOs, particularly PBHs, arise from the GW background generated by their formation mechanism in the early Universe, such as the collapse of large primordial density fluctuations~\cite{NANOGrav:2023hfp}. DSs, in contrast, form much later, as detailed in Sec.~\ref{sec:formation}. Therefore, PTAs do not offer direct constraints on the DSs energy fraction, leaving microlensing as the primary observational constraint.

In Fig.~\ref{fig:detectability},  regions with BDS merger rate $R > 1\,{\rm Gpc^{-1}\, yr^{-1}}$ at redshifts $z = 0.1,\,1,\,5$ are illustrated in gray. The regions that, in addition, comply with the criterion of Eq.~(\ref{SNR8}) are depicted by blue (LISA) and red (DECIGO). Therefore within a year of observations, it is evident that both LISA and DECIGO will be able to probe a large parameter space of DSs. The fact that detectable mergers of BDS can be observed for various redshifts suggests that the BDS merger rate redshift dependence as shown in Fig.~\ref{fig:mergerrate_redshift} can be tested.
Regions ruled out by microlensing observations (see Eq.~\eqref{eq:microlensing}) are depicted in cyan.
In the upper panels of Fig.~\ref{fig:detectability},  
we present the parameter space that can be probed by LISA and DECIGO in terms of the underlying DM model parameters i.e., the DM and dark photon masses for a fixed $\alpha_D=0.1$. One can see that the experiments 
 can probe DSs at the  upper left and  lower right corners of the plots. The former probes DM masses roughly between 10 keV and a few GeV with dark photon mass ranging between $\sim100$ eV and $\sim 100$ keV, while the latter probes DM masses higher than a 100 GeV and a dark photon mass ranging from eV to keV.

% parameter regime under the microlensing constraint, which corresponds with $(m_{e_D}, m_{\gamma_D}) \sim (1, 10^{-5}) \,{\rm GeV}$ and $(100, 10^{-9}) \, {\rm GeV}$. 

In the lower panels of Fig.~\ref{fig:detectability}, we have chosen to present the DS parameter space that can be probed in terms of star properties i.e., the DS mass and compactness. The plots show that LISA and DECIGO can detect DSs with  $M_{\rm DS} > 10^2 \, M_\odot$ and $C>10^{-3}$ (which are not excluded by the microlensing constraints). As we mentioned before, the parameter $f_{e_D}$ is subjected to constraints from the Bullet Cluster and the ellipticity of galaxies and it should not surpass $\sim 10\%$ of the DM abundance. To be more precise it is the free dark electrons (not those bound in DSs) that are subjected to the constraint, i.e., the combination $f_{e_D}(1-f_{e_D-\rm DS})$ should be no more than $\sim 0.1$. We have chosen $f_{e_D-\rm DS}=0.5$. Note that for baryons in the Milky Way this is of the same ballpark i.e., 0.85.
On the other hand the fraction of DSs $f_{\rm DS}=f_{e_D}f_{e_D-\rm DS}$ is subjected to the MACHO microlensing constraints we have mentioned. In Fig.~\ref{fig:detectability}, we present the parameter space ruled out by these observations using the cyan-colored region for the choices of $f_{e_D}=0.1,0.05$ for the left and right panels, respectively. We note that these constraints mainly rule out models for DS masses smaller than $\sim100M_\odot$.  However both LISA and DECIGO, well-suited for observation of mergers of very massive objects, would still be able to observe signals for $M>100M_\odot$. Since $f_{\rm DS} \propto f_{e_D}$ and $R_{\rm sp} \propto f_{\rm DS}^2 \propto f_{e_D}^2$, the decrease in the value of $f_{e_D}$ would effectively shrink both the detectable  and the microlensing constraint regions and move the microlensing constraint away from the detectable region. When $f_{e_D} < 0.023$, the corresponding microlensing observations do not constrain DS merger regions with $R > 1\,{\rm Gpc^{-1}\, yr^{-1}}$ any more.

\begin{figure}[t!]
	\centering
	\includegraphics[width=.95\textwidth]{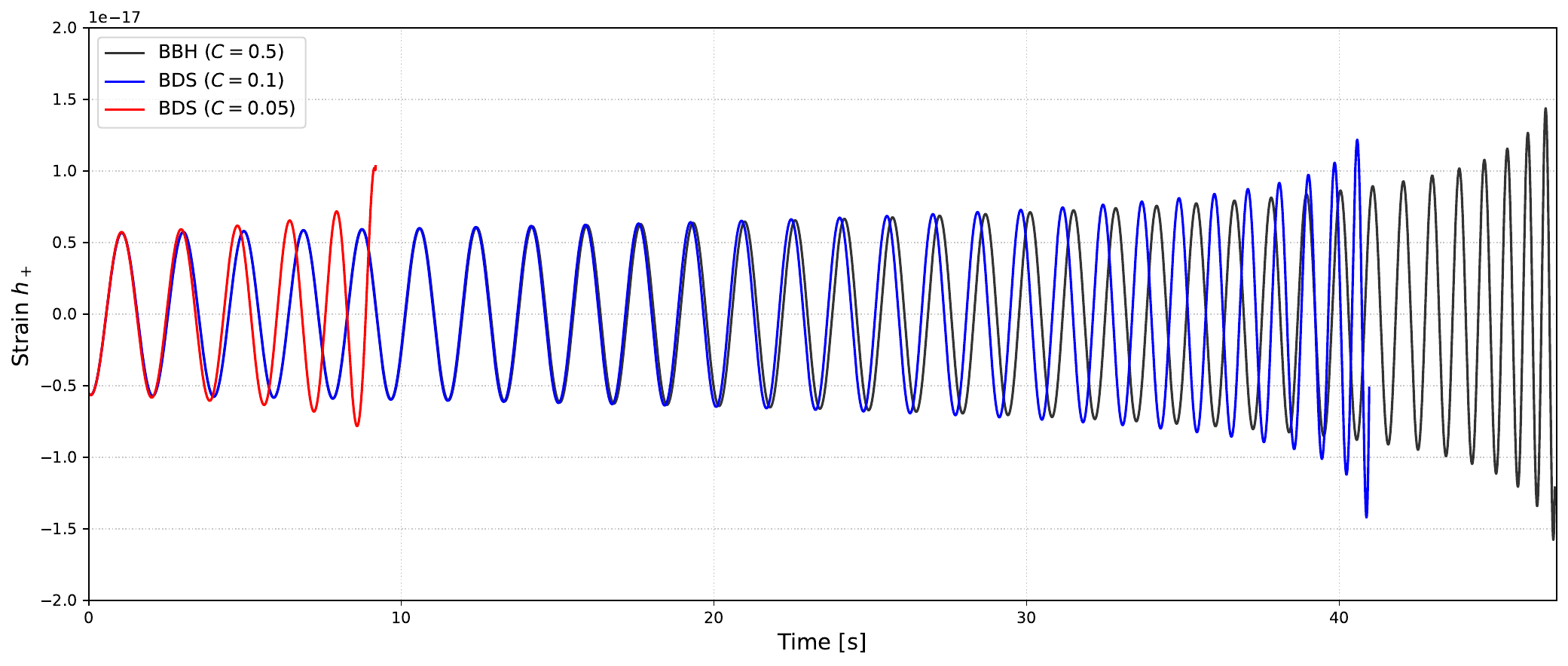}
	\caption{The GW waveform of BDS mergers with benchmark parameters of $(m_{e_D}, m_{\gamma_D}) = (2.18, 6\times10^{-5}) \, \text{GeV}, \,(1.91, 4\times10^{-5}) \, \text{GeV}$ which correspond to a DS mass of $1000\, M_\odot$ and DS compactness $C = 0.1,~ 0.05$, denoted as blue and red curves respectively. The same BH mass binary is also illustrated as  black curve to highlight the difference of GW waveforms between BBH and BDS events.}
	\label{fig:gw_waveform}
\end{figure}

For a complete picture of detectability of BDS mergers, we also show the expected GW waveform of BDS events in comparison  to  the GW waveform of a BBH event with the same masses. To numerically calculate the GW waveform of the BDS merger, we use the public code \textsc{PyCBC} package and employ a time-domain TaylorT4 approximant to obtain the inspiral phase GW waveforms~\cite{pycbc_software}. The main difference in GW waveforms between BDSs and BBHs comes from the tidal deformability. The tidal deformation effect on the GW waveform of BDSs is determined by the dimensionless tidal deformability parameter $\Lambda_2$. This parameter $\Lambda_2$ is related to the compactness of DS as~\cite{Damour:2009vw}
\begin{equation}
    \Lambda_2 = \frac{2}{3} k_2 C^{-5}~,
\end{equation}
where $k_2$ is the tidal Love number of DS. We have a detailed discussion on how to estimate  $k_2$  in Appendix~\ref{app:love2}. By using compactness $C=0.1$ and $0.05$ with DS mass of $1000\,M_\odot$ in the  \textsc{PyCBC} code, we generate the corresponding GW waveform as shown in Fig.~\ref{fig:gw_waveform}. 
These two GW waveforms of BDSs correspond to the benchmark parameter values $(m_{e_D}, m_{\gamma_D}) = (2.18, 6\times10^{-5}) \, \text{GeV}$ and $(1.91, 4\times10^{-5}) \, \text{GeV}$ labelled as stars in Fig.~\ref{fig:detectability}. When comparing them with the GW waveform of the BBH event (black), it clearly shows that the GW waveforms of BDSs agree with that of BBH at low GW frequency. However at higher frequencies  tidal deformation  dominates the waveform evolution as the BDSs approach the merger phase. Another feature in GW waveforms of BDSs is their maximum frequency in the inspiral phase is lower in smaller compactness BDS mergers than  in larger. This can be intuitively understood as a low compactness BDS system can be tidally disrupted more easily and enter into the merger phase earlier. 

%Some qualitative features of Fig.~\ref{fig:gw_waveform} can be easily extracted along the lines of~\cite{Giudice:2016zpa}. From Kepler's third law the orbital period of a binary system is 
%\begin{equation}
 %   P^2=\frac{4\pi^2 l^3}{G M_{\rm tot}},
%\end{equation}
%where $l$ is the semi-majot axis of the elliptical orbit, and $M_{\rm tot}$ the total mass of the binary. Given that the GW frequency is $2/P$ (twice the orbital one), one can estimate the frequency at which the inspiral phase is over for a BBH by taking the sum of the two innermost stable circular orbit (ISCO) radii $R_{\rm BH}^{\rm ISCO}=6 GM_{\rm tot}$. This gives the frequency where the inspiral phase is over as
%\begin{equation}
%    f_{\rm BH}^{\rm ISCO}=\frac{1}{6^{3/2}\pi GM_{\rm tot}}. 
%\end{equation}
%For a DS binary with each DS having a compactness $C=M/R$, the end of the inspiral phase should scale 

\section{Conclusions}\label{sec:conclusion}

In this work, we have studied the formation and structure of asymmetric DM stars in a dissipative dark-sector model and calculated their GW signatures, with particular emphasis on binary mergers occurring inside DM spikes around SMBHs. We have also identified the regions of parameter space that can yield a large number of observable GW events and that place the signals within the sensitivity range of the future GW telescopes LISA and DECIGO.

We find that the compactness of the DS strongly affects the binary merger rate: for moderate compactness, $10^{-5}\lesssim C\lesssim 10^{-2}$, tidal energy deposition during close encounters enhances the merger cross section relative to the BH limit, while for very diffuse objects direct collisions dominate, but tend to produce signals largely outside the current detector bands. Both LISA and DECIGO should be able to detect DS mergers with $\mathrm{SNR}\geq 8$ in a one-year observation for masses and compactness roughly satisfying $M_{\rm DS}\gtrsim 10^2\,M_\odot$ and $C\gtrsim 10^{-3}$, assuming DM parameters compatible with microlensing constraints. We have further estimated the redshift evolution of the DS merger rate by considering their population inside dense DM spikes surrounding SMBHs. Using the SMBH mass function and imposing that DSs form only within a specific range of host halo masses, we computed the merger rate evolution with redshift. For our benchmark scenarios, consisting of DSs with mass $1000\,M_\odot$ and compactness as large as $C=0.1$, the resulting merger rate can exceed that of Population~I/II BHs by a factor of $\sim 5$ at low redshift. We have also examined the dependence of this result on both the DS compactness and the spike power-law index $\gamma$: in general, smaller compactness and smaller $\gamma$ reduce the merger rate, which may render it subdominant with respect to Pop~I/II BHs. Nevertheless, even in the most conservative case, the predicted rate remains large enough to yield observable signals. Finally, DS mergers can be distinguished from BH mergers through their GW signatures. Although DS waveforms closely track those of BH systems during the low-frequency inspiral, they exhibit measurable deviations at higher frequencies, especially during the late inspiral and near-merger phases, due to finite-size and tidal effects. Given the mass and compactness of the DSs considered here, it is difficult to identify ordinary astrophysical formation channels that would mimic the same behavior. Therefore, the detection of such deviations would provide strong evidence for the existence of asymmetric DSs.

\acknowledgments

The work of BBK and QD is supported by IBS under the project code IBS-R018-D3.

\clearpage
\appendix
\section*{Appendix}
\section{Tidal energy loss}\label{app:tidal_loss}
The computation of tidal energy loss during close encounters follows the methodology established by Press and Teukolsky~\citep{1977ApJ...213..183P}. For a star of mass $M_i$ and radius $R_i$ perturbed by a companion of mass $M_j$, the energy deposited into stellar oscillations during a parabolic encounter is given by:

\begin{equation}
\Delta E_i = \frac{GM_j^2}{R_i}  \sum_{l=2,3,\dots} \left( \frac{R_i}{R_p} \right)^{2l+2} T_{\ell,i}(\eta_i),
\end{equation}
where $R_p$ is the periastron distance and the dimensionless parameter $\eta_i$ is defined as:
\begin{equation}
\eta_i \equiv \left( \frac{M_i}{M_i + M_j} \right)^{1/2} \left( \frac{R_p}{R_i} \right)^{3/2}.
\end{equation}

The function $T_{\ell,i}(\eta)$ encapsulates the efficiency of energy transfer through mode excitation and is computed as:
\begin{equation}
T_{\ell,i}(\eta) \equiv 2\pi^2 \sum_n |Q_{nl}^{i}|^2 \sum_{m=-l}^{l} |K_{nlm}^{i}|^2,
\end{equation}
where the sums run over the discrete set of stellar eigenmodes (indexed by $n$) and the azimuthal index $m$. The overlap integral $Q_{nl}$ measures the coupling between the tidal potential and the stellar normal modes:

\begin{equation}\label{eq:Q_eq}
Q^i_{nl} \equiv (\rho_{c,i}/(M_i/R_i^3))\int_0^1 d\hat{r}\,\hat{r}^{\ell+1}  \hat{\rho}_i [\xi^R_{nlm} + (l+1)\xi^S_{nlm}]
\end{equation}
where $\xi^R_{nlm}$ and $\xi^S_{nlm}$ are the radial and poloidal components of the eigenfunctions, expressed in natural units ($\hat{r} = r/R_i$, $\bar{\rho}_i = \rho_i / \rho_{c,i}$), and normalized such that $\int \xi_{nlm} \cdot \xi^*_{n'l'm'} d^3x = \delta_{nn'}\delta_{ll'}\delta_{mm'}$.

The temporal coupling integral $K_{nlm}$ accounts for the orbital dynamics:

\begin{equation}
K^{i}_{nlm}(\eta) \equiv \frac{W_{lm}}{2\pi}2^{3/2}\eta I_{\ell m}(\eta\Omega^i_n)
\end{equation}
with the coefficient $W_{lm}$ given by:

\begin{equation}
W_{lm} \equiv (-)^{(l+m)/2}\frac{\sqrt{\frac{4\pi}{2l+1} (l-m)! (l+m)!}}{2^l \left( \frac{l-m}{2} \right) ! \left( \frac{l+m}{2} \right) !}, 
\end{equation}
where $(-)^k$ is interpreted as zero if $k$ is not an integer. The functions $I_{\ell m}$ are defined as
\begin{equation*}
    I_{\ell m}(y)=\int\displaylimits_0^\infty dx(1+x^2)^{-\ell}\cos\left[\sqrt{2}y(x+x^3/3)+2m\arctan(x)\right],
\end{equation*}
but for practical computations, we use the approximations obtained by~\cite{1977ApJ...213..183P} in section VI of the original work.

In practice, only the quadrupole ($l=2$) and octupole ($l=3$) terms are considered, as higher-order contributions are negligible. The eigenfunctions and eigenfrequencies are computed for the specific stellar model.

The  eigenvalue problem for the normal modes of a spherically-symmetric star is given by~\cite{Pani:2014rca,1980tsp..book.....C,1971AcA....21..289D}
\begin{align}
    \frac{d\hat{y}_1}{d\hat{r}}&=-\frac{2}{\hat{r}}\hat{y}_1+\left(\frac{\hat{y}_2}{\hat{\kappa}(\hat{r})}\right)+\frac{\ell(\ell+1)}{\hat{r}}\hat{y}_3,\cr
    \frac{d\hat{y}_2}{d\hat{r}}&=\alpha_1\left\{-\left[\Omega^2\hat{\rho}(\hat{r})+4\frac{\hat{g}(\hat{r})\hat{\rho}(\hat{r})}{\hat{r}}\right]\hat{y}_1+\frac{\ell(\ell+1)\hat{g}(\hat{r})\hat{\rho}(\hat{r})}{\hat{r}}\hat{y}_3-\hat{\rho}(\hat{r}) y_6\right\},\cr
    \hat{y}_3&=\frac{1}{\hat{r}\Omega^2}\left(\hat{g}(\hat{r})\hat{y}_1-\frac{\hat{y}_{2}}{\alpha_1\hat{\rho}(\hat{r})}-\hat{y}_5\right),\cr
    \frac{d\hat{y}_5}{d\hat{r}}&=\alpha_2\hat{\rho}(\hat{r}) \hat{y}_1+\hat{y}_6,\cr
    \frac{d\hat{y}_6}{d\hat{r}}&=-\alpha_2\frac{\ell(\ell+1)\hat{\rho}(\hat{r})}{\hat{r}}\hat{y}_3+\frac{\ell(\ell+1)}{\hat{r}^2}\hat{y}_5-\frac{2}{\hat{r}}\hat{y}_6,
\end{align}
where we work in terms of the dimensionless variables $\hat{y_1}=y_1/R$, $\hat{y_2}=y_2/P_c$, $\hat{y}_3=y_3/R$, $\hat{y_5}=y_5/(GM/R)$ and $\hat{y}_6=y_6/(GM/R^2)$. The functions $\hat{g}(\hat{r})=\hat{m}(\hat{r})/\hat{r}^2$ and $\hat{\kappa}(\hat{r})=\hat{\rho}d\hat{P}/d\hat{\rho}$ are the dimensionless gravitational acceleration and incompressibility, respectively. Here, we also consider the normalized quantities $\hat{r}\equiv r/R$, $\hat{m}\equiv m/M$, $\hat{\rho}\equiv \rho/\rho_c$ and $\hat{P}=P/P_c$. 
We also define the constants $\alpha_1=GM\rho_c/(P_c R)$, $\alpha_2=4\pi\rho_c R^3/M$ and the dimensionless eigenfrequency squared $\Omega^2=R^3\omega^2/(GM)$, where $\omega$ is the dimensionful value.
We impose regularity at the center $\hat{r}=0$, which leads to the first-order behavior of the solutions:
\begin{align}
    \hat{y}_1&=\hat{A}\hat{r}^{\ell-1},\cr
    \hat{y}_2&=\frac{\alpha_1}{3\ell}\left[\alpha_2\ell\hat{A}-3\ell\hat{B}-3\Omega^2\hat{A}\right]\hat{r}^\ell,\cr
    \hat{y}_3&=\frac{\hat{A}}{\ell}\hat{r}^{\ell-1},\cr
    \hat{y}_5&=\hat{B}\hat{r}^{\ell},\cr
    \hat{y}_6&=\left(\ell\hat{B}-\alpha_2\hat{A}\right)\hat{r}^{\ell-1},
\end{align}
where $\hat{A}$ and $\hat{B}$ are dimensionless constants to be determined by imposing the boundary conditions at the surface:
\begin{equation*}
    \hat{y}_2(\hat{r}=1)=0,\hspace{0.5cm}\hat{y}_6(\hat{r}=1)+(\ell+1)\hat{y}_5(\hat{r}=1)=0.
\end{equation*}
The dimensionless functions $\hat{y}_i(\hat r)$ and the physical displacement components are related as
\begin{equation}
\xi^R_{n\ell}(r) = R_i\,\hat y_1(\hat r), \qquad
\xi^S_{n\ell}(r) = R_i\,\hat y_3(\hat r).
\end{equation}
Thus $\hat y_1$ and $\hat y_3$ are the radial and poloidal eigenfunctions expressed in the same nondimensional units adopted above.
Therefore, the dimensionless overlap integral~\eqref{eq:Q_eq} is, in this notation,
\begin{equation}\label{eq:g_raw}
Q^i_{n\ell} = \sqrt{(\rho_{c,i}/(M_i/R_i^3))}\frac{\int_0^1 d\hat r\, \hat\rho_i(\hat r)\hat{r}^{\ell+1}
\left[\hat{y}_1(\hat {r}) + (\ell+1)\hat {y}_3(\hat{r})\right]}{\int_0^1 d\hat {r}\,\hat{r}^2 \hat{\rho}_i(\hat{r})\left[\hat{y}_1(\hat{r})^2 + \ell(\ell+1)\hat{y}_3(\hat{r})^2\right]},
\end{equation}
where the denominator accounts for the normalization $\int \xi_{nlm} \cdot \xi^*_{n'l'm'} d^3x = \delta_{nn'}\delta_{ll'}\delta_{mm'}$.

\section{Calculation of the $\ell=2$ tidal Love number}
\label{app:love2}

When a star is placed in an external tidal field, it develops a quadrupole moment in response.  For a static, quadrupolar tidal field $\mathcal{E}_{ij}$, the induced quadrupole moment $Q_{ij}$ of the star is proportional to the applied field:
\begin{equation}
Q_{ij} = -\Lambda_2\,\mathcal{E}_{ij},
\end{equation}
where $\Lambda_2$ is the tidal deformability.  The dimensionless Love number $k_2$ is defined by
\begin{equation}
k_2 \equiv \frac{3}{2}\,\frac{\Lambda_2}{R^5},
\end{equation}
with $R$ the stellar radius.  Equivalently, $k_2$ can be extracted from the asymptotic behaviour of the metric perturbation far from the star and it encodes the star's internal structure and its response to tidal forces.

\subsection{Interior perturbation equation}

Inside the star we work in the Regge–Wheeler gauge and consider static, even‑parity perturbations.  The perturbation variable $H(r)$ (which describes the distortion of the metric) satisfies a second‑order ordinary differential equation \cite{Damour:2009vw}:
\begin{equation}
\label{eq:H_ell_in}
\frac{d^2H_\ell}{dr^2}+A(r)\frac{dH_\ell}{dr}+Q_\ell(r)H_\ell=0,
\end{equation}
where $\ell$ is the multipole index.  The coefficients are
\begin{align}
A(r) &\equiv \frac{2}{r}+\mathrm{e}^{\lambda(r)}\!\left[\frac{2Gm(r)}{r^2}+4\pi G r\bigl(P(r)-\rho(r)\bigr)\right],\\[4pt]
Q_\ell(r) &\equiv \mathrm{e}^{\lambda(r)}\!\left[-\frac{\ell(\ell+1)}{r^2}+4\pi G\left(5\rho(r)+9P(r)+\frac{\rho(r)+P(r)}{(\partial P/\partial \rho)(r)}\right)\right]
-\bigl(\nu'(r)\bigr)^2,\\[4pt]
\mathrm{e}^{\lambda(r)} &\equiv \left(1-\frac{2Gm(r)}{r}\right)^{-1},\qquad 
\nu'(r) \equiv 2G\mathrm{e}^{\lambda(r)}\frac{m(r)+4\pi r^3 P(r)}{r^2}.
\end{align}
Here $m(r)$ is the mass inside radius $r$, $\rho(r)$ the energy density, and $P(r)$ the pressure.  The function $\nu(r)$ is the other metric potential, related to $g_{tt}=\mathrm{e}^{2\nu}$.

It is convenient to introduce the dimensionless function
\begin{equation}
y_\ell(r) \equiv \frac{r H_\ell'(r)}{H_\ell(r)} .
\end{equation}
From Eq.~\eqref{eq:H_ell_in} one derives a first‑order Riccati equation for $y_\ell$:
\begin{equation}
r\frac{dy_\ell}{dr}+y_\ell^2(r)+y_\ell(r)\,\mathrm{e}^{\lambda(r)}\!\left[1+4\pi G r^2\bigl(P(r)-\rho(r)\bigr)\right]+r^2Q_\ell(r)=0.
\label{eq:y_riccati}
\end{equation}
Near the centre $r=0$ regularity of the perturbation requires
\begin{equation}
H_\ell(r) = a r^\ell\bigl[1+\mathcal{O}(r^2)\bigr],
\end{equation}
which gives the boundary condition
\begin{equation}
y_\ell(0)=\ell .
\end{equation}
For $\ell=2$ we therefore have $y_2(0)=2$.

\subsection{Exterior solution}

Outside the star ($r>R$) the energy density and pressure vanish, $m(r)=M$, and $\mathrm{e}^{\lambda(r)}=(1-2GM/r)^{-1}$.  Eq. \eqref{eq:H_ell_in} reduces to an associated Legendre equation.  Setting $x = r/M-1$ ($x>1$ for $r>R$), the general solution for $\ell=2$ is
\begin{equation}
H_2(x) = c_1 Q_2^2(x) + c_2 P_2^2(x),
\end{equation}
where
\begin{align}
P_2^2(x) &= 3(x^2-1),\\[4pt]
Q_2^2(x) &= \frac{3}{2}(x^2-1)\ln\!\left(\frac{x+1}{x-1}\right) + \frac{x(5-3x^2)}{x^2-1}.
\end{align}
Returning to the compactness variable $C_r \equiv G M/r$ (note that at the surface $C_R = C = G M/R$), we can rewrite the exterior solution as
\begin{equation}
\label{eq:H2_ext}
H_2(C_r)=\frac{1-2C_r}{C_r^{\,2}}\Biggl\{
c_1\!\left[
\frac{C_r(1-C_r)(2C_r^2+6C_r-3)}{(1-2C_r)^2}
-\frac{3}{2}\ln(1-2C_r)
\right]
+ 3c_2
\Biggr\}.
\end{equation}

At large distance ($C_r\to0$) the expansion of \eqref{eq:H2_ext} yields
\begin{equation}
H_2 \to c_1\frac{8}{5}C_r^3 + c_2\frac{3}{C_r^2} + \mathcal{O}(C_r^{-1}).
\end{equation}
The term $\propto C_r^{-2}$ corresponds to the external quadrupolar tidal field, while the term $\propto C_r^3$ gives the induced quadrupole moment.  Matching to the metric perturbations in the asymptotically flat region \cite{Damour:2009vw} identifies the constants
\begin{equation}
c_2 = \frac{1}{3}\mathcal{E} G^2M^2,\qquad 
c_1 = \frac{15\Lambda_2\mathcal{E}}{8G^3M^3},
\end{equation}
where $\mathcal{E}$ is the tidal field strength.  Using $\Lambda_2 = \frac{2}{3}k_2 R^5$ and $C = G M/R$, we can express $c_1$ in terms of $k_2$ and $C$:
\begin{equation}
c_1 = \frac{5 R^2 k_2\mathcal{E}}{4C^3}.
\end{equation}

\subsection{Surface matching and the Love number formula}

\begin{figure}[t!]
	\centering
	\includegraphics[width=.58\textwidth]{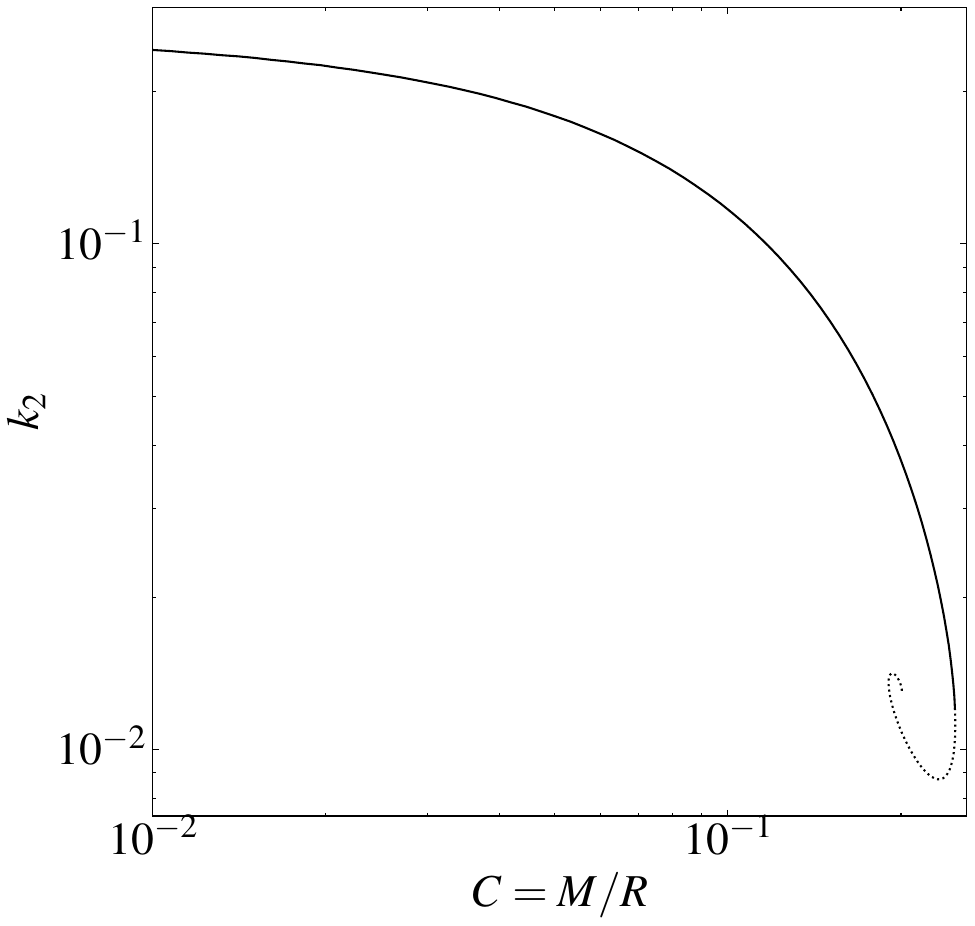}
	\caption{Love number $k_2$ as a function of the compactness for a DS with EoS given by Eq.~\eqref{eq:EOS}.}
	\label{fig:LoveNumbers}
\end{figure}

The interior solution provides the value $y_R \equiv y_2(R)$ at the stellar surface.  From the exterior solution \eqref{eq:H2_ext} we compute $y_R$ as
\begin{equation}
y_R = -C \frac{dH_2/dC}{H_2}\bigg|_{C_r=C}.
\end{equation}
Performing the derivative and simplifying gives an algebraic relation among $y_R$, $C$, and $k_2$.  Solving for $k_2$ yields the well‑known expression \cite{Damour:2009vw}:
\begin{align}
\label{eq:k2_final}
k_2 =&\ \frac{8}{5}C^5(1-2C)^2\bigl[2+2C(y_R-1)-y_R\bigr] \nonumber\\
&\times\Bigg\{2C\Bigl[6-3y_R + C\bigl(3(5y_R-8) \nonumber\\
&\qquad + 2C\bigl(13-11y_R + C(-2+3y_R+2C(1+y_R))\bigr)\bigr)\Bigr] \nonumber\\
&\quad +3(1-2C)^2\bigl[2+2C(y_R-1)-y_R\bigr]\ln(1-2C)\Bigg\}^{-1}.
\end{align}
Equation \eqref{eq:k2_final} determines the Love number $k_2$ solely from the compactness $C$ and the value of $y$ at the surface.  The quantity $y_R$ must be obtained by integrating the interior Riccati equation~\eqref{eq:y_riccati} from the center (with $y_2(0)=2$) out to $r=R$. In Fig.~\ref{fig:LoveNumbers}, we present the $k_2$ Love number as a function of compactness for the DSs we consider in this work. As in Fig.~\ref{fig:mass_vs_r}, the solid branch corresponds to stable configurations, while the dashed one to unstable solutions.

\bibliographystyle{JHEP} % We choose the "plain" reference style
\bibliography{Bibliography} % Entries are in the References.bib file

\end{document}